\documentclass[%
  twocolumn
]{mpi2015-cscpreprint}

\usepackage[american]{babel}
\usepackage{placeins}
\usepackage{nomencl}
\usepackage{graphicx}
\usepackage[caption=false]{subfig}
\usepackage{amssymb}
\usepackage{amsthm}
\usepackage{amsmath}
\usepackage{verbatim}
\begin{document}
  
%%%%%%%%%%%%%%%%%%%%%%%%%%%%%%%%%%%%%%%%%%%%%%%%%%%%%%%%%%%%%%%%%%%%%%%%%%%%%%%%
% PAPER INFORMATION.                                                           %
%%%%%%%%%%%%%%%%%%%%%%%%%%%%%%%%%%%%%%%%%%%%%%%%%%%%%%%%%%%%%%%%%%%%%%%%%%%%%%%%

\title{Adjacency-based, non-intrusive model reduction for Vortex-Induced Vibrations}

\author[$1$,$\ast$]{Leonidas Gkimisis}
\author[$2$]{Thomas Richter}
\author[$1,2$]{Peter Benner}

\affil[$1$]{Max Planck Institute for Dynamics of Complex Techincal Systems, Sandtorstra{\ss}e 1, 39106 Magdeburg, Germany,
Computational Methods in Systems and Control Theory (CSC).}

\affil[$2$]{Otto-von-Guericke Otto-von-Guericke-
Universit\"at Magdeburg, Universit\"atsplatz 2, 39106 Magdeburg, Germany, Faculty of Mathematics, Institute for Analysis and Numerics.}

\affil[$\ast$]{corresponding author, \email{gkimisis@mpi-magdeburg.mpg.de}}

% \authorcr
%   \email{thomas.richter@ovgu.de}, \orcid{0000-0001-7721-0426}}

% \affil[$3$]{Max Planck Institute for Dynamics of Complex Techincal Systems, Sandtorstrasse 1, 39106 Magdeburg, Germany,
% Computational Methods in Systems and Control Theory (CSC).}

% \affil[$3$]{Otto-von-Guericke University Magdeburg, University Square 2, 39106 Magdeburg, Germany, Faculty of Mathematics, Institute for Analysis and Numerics.\authorcr
%   \email{benner@mpi-magdeburg.mpg.de}, \orcid{0000-0003-3362-4103}}
  
\shorttitle{Adjacency-based, non-intrusive model reduction for VIV}
\shortauthor{L. Gkimisis, T. Richter, P. Benner}
\shortdate{}
  

  
\abstract{%
Vortex-induced vibrations (VIV) pose computationally expensive problems of high
practical interest to several engineering fields. In this work we develop a non-intrusive,
reduced-order modelling  methodology for two-dimensional (2D) VIV simulations. We consider an elliptical,
non-deformable solid mounted on springs, subject to a laminar, incompressible flow.
Approximating the Arbitrary Lagrangian-Eulerian (ALE) incompressible
Navier-Stokes (NS) formulation, a discrete-time, quadratic-bilinear data-driven model structure is assigned for the velocity flowfield prediction. The full-order, data-driven model operators have a predefined sparsity pattern, following the adjacency-based sparsity of the discretized ALE-NS operators. Thus, the data-driven operators inference requires solving many low-dimensional least squares problems, isolating the contribution of ``nearest neighbours" for each degree of freedom. Numerical aspects such as data centering and regularization are extensively
discussed. With this approach, Dirichlet boundary conditions at the inlet
and the fluid/solid interface can be enforced on the full-order, non-intrusive level. Consequently, a non-intrusive reduced-order model (ROM) for the velocity flowfield is obtained after linear projection. The resulting data-driven ROM is coupled with the first-principle, 2D solid oscillation equations and is simulated using an implicit time integration scheme. The coupled solution is then mapped on the deformed grid through the inverse ALE map. This methodology is showcased for two
testcases, at different Reynolds numbers ($Re=90,180$). Numerical results indicate a successful coupling
between the data-driven velocity flowfield and the solid oscillation, with prediction errors of less than $3 \%$ for both the flowfield and the solid oscillation. A comparative study with respect to the ROM dimension indicates the robustness and potential of the approach.
}

\novelty{Non-intrusive reduced-order modeling method for VIV problems. Sparse, nonlinear data-driven model inference. Coupled data-driven flow ROM with first principle 2D oscillation dynamics.}

\maketitle
  
\section{Introduction}%
\label{sec:intro}

Vortex-Induced vibrations (VIV) comprise a class of Fluid-Structure Interaction (FSI) problems with high practical interest to numerous engineering fields, among which are wind, offshore and aerospace engineering \cite{Sarpkaya2004, Williamson2004}. Vortex-induced vibrations concern the two-way coupled system response of a non-deformable solid body mounted on elastic supports, subject to a fluid flow. A multitude of complex dynamical phenomena arise from the interplay between solid and fluid dynamics, through the mechanism of asymmetric vortex shedding in the wake of the solid body \cite{Bearman2009}. Vortex shedding patterns, as well as oscillation coupling properties can greatly vary, depending on the flow conditions and solid parameters of a given configuration \cite{Kang2021, Govardhan2000}. 

In practice, the dynamically varying loads resulting from vortex shedding can lead to strong vibrations and thus a significant decrease in the fatigue life of structures. Such vibrations could be proven particularly destructive in applications such as underwater pipelines \cite{Kang2020}, turbine poles \cite{Carlson2018} or structural cables \cite{Jafari2020} and are thus carefully accounted for during engineering design. Apart from VIV suppression, intensive research is being performed on efficient harnessing of the kinetic energy of vortex-induced vibrations for energy production \cite{Zheng2020,Grouthier2014}. Both these fields indicate the strong motivation in developing efficient VIV simulation tools for engineering design optimization and control.

Over the past decades, a number of methods have been developed for the direct numerical simulation of FSI and VIV problems \cite{Takizawa2012,Fernandez2007,Bazilevs2013}, confronting issues such as numerical stability and computational overhead \cite{Richter2017,Lozovskiy2019}. Concise reviews of such numerical methods for FSI problems are given in \cite{He2018,Dowell2001}. Nonetheless, even for regimes of incompressible, laminar flows, the two-way coupling of two high-dimensional subsystems (i.e. a CFD and a structural finite element model) remains a challenging task. Depending on the problem at hand and the employed numerical solution, stability is not always guaranteed \cite{Revstedt}, while the stiff coupling and high dimensionality introduce a significant computational cost \cite{Richter2017}. 

The growing field of model order reduction (MOR) comprises a potential alternative to bridge the gap between accuracy and computational overhead. MOR methods have proven to be efficient for several demanding, high-dimensional fluid and solid dynamics applications, leveraging computational cost and accuracy \cite{Cai2021,Schmid2011,Peherstorfer2016,Gosea2022}. A number of research works have also been presented for different FSI aspects and applications. Notably, Liberge et al. employed the method of Proper Orthogonal Decomposition (POD) to the flow around an oscillating cylinder, transferring the FSI problem to a global, fixed domain \cite{Liberge2010}. Using experimental Particle Image Velocimetry (PIV) data, Riches et al. also employed POD, to investigate the intrinsic mechanisms of VIV wake-dynamics. A reduced basis method differentiating between solid, fluid and deformation modes for intrusive model reduction was developed by Nonino et al., applied to both partitioned and monolithic FSI numerical models \cite{Nonino2021}. Focusing on parameter-dependent problems, Benner et al. proposed a low-rank method to increase the Newton iteration's computational efficiency for FSI computations \cite{BenFSI}, while Lieu et al. produced a POD-based model for the aeroelastic response of a complete F-16 aircraft, interpolating the model response over varying free-stream Mach numbers \cite{Lieu2005}. 

In cases where only simulation or experimental data are available, non-intrusive model reduction methods are being employed for modeling and prediction purposes. Working in this direction, Poussot-Vassal et al. \cite{Poussot-Vassal2018} used the Loewner framework for the prediction of aircraft response to gust forcing. Focusing on the aeroelastic phenomenon of stall flutter, Dai et al. developed a Recurrent Neural Network architecture for the non-intrusive prediction of pitching airfoil dynamics \cite{dai_2023_stall}. In a more general framework, Xiao et al. \cite{Xiao2016} presented a non-intrusive reduced-order model for FSI problems, based on radial basis function interpolation over time. In a slightly different direction, Yao et al. \cite{Yao2017} focused on VIV problems at low Reynolds numbers, proposing a linear, Eigenvector Realization Algorithm (ERA) approach to explore the mechanism of transition to vortex shedding for different VIV configurations. In this study, we focus on laminar, low mass ratio, 2-dimensional VIV testcases \cite{Navrose2014,Prasanth2009}, as a prototype example for the proposed non-intrusive, ROM methodology.

In further detail, we present a methodology for data-driven model reduction, focusing on the prediction of 2D laminar, vortex-induced vibration dynamics. Assuming state access, a non-intrusive model for the incompressible fluid flow is constructed on an ALE reference grid. To account for the grid deformation, the velocity data being interpolated to a reference domain through an ALE map, given by the solution of a Laplace equation. We show that the structure of the ALE-NS formulation for the velocity flowfield is approximately quadratic-bilinear. Motivated by the adjacency-based sparsity of the discretized ALE-NS operators, we propose a novel method to construct a full-order, data-driven model for the velocity flowfield, with an a priori adjacency-based, sparsity pattern. As a result, we locally infer the data-driven operators by solving one least squares problem for each internal grid node with an $L_2$ regularization term. In parallel, a proper treatment of the boundary nodes allows for a direct enforcement of the Dirichlet boundary conditions at the fluid inlet and the fluid/solid interface on the full-order level. The data-driven fluid subsystem is then coupled with the first-principle solid motion oscillations through a time-implicit scheme. The map to the deformed configuration is also considered as a post-processing step of the VIV predictions. The proposed methodology is showcased for two VIV cases of an oscillating solid along the streamwise and transverse directions, subject to a $Re=90$ and $Re=180$ flow. Through these testcases, several properties as well as the predictive capabilities of the method are highlighted.

The rest of this work is structured in the following way: A review of the theoretical background on vortex-induced vibrations is given in \Cref{sec:the}, motivating the non-intrusive model formulation for the velocity flowfield. The developed methodology is analyzed in \Cref{sec:meth}, for the construction of a non-intrusive fluid dynamics ROM and its coupling with the solid oscillations along the transverse and streamwise directions. Finally, the model is tested for two testcases with $Re=90$ and $Re=180$ flow past an ellipse-shaped body, with corresponding results given in \Cref{sec:res}. Conclusions and potential future work is discussed in \Cref{sec:con}.

%%%%%%%%%%%%%%%%%%%%%%%%%%%%%%%%%%%%%%%%%%%%%%%%%%%%%%%%%%%%%%%%%%%%%%%%%%%%%%%%

\section{Theoretical Background}%
\label{sec:the}

In this section, we present the physical modeling of laminar, vortex-induced vibration problems, which motivates the structure of the developed non-intrusive, data-driven model for the fluid velocity field. 

%%%%%%%%%%%%%%%%%%%%%%%%%%%%%%%%%%%%%%%%%%%%%%%%%%%%%%%%%%%%%%%%%%%%%%%%%%%%%%%%

\subsection{Vortex-Induced Vibrations}%
\label{subs:eqns}

\begin{figure}[!htbp]
  \begin{center}
    \resizebox{8cm}{!}{
    \includegraphics{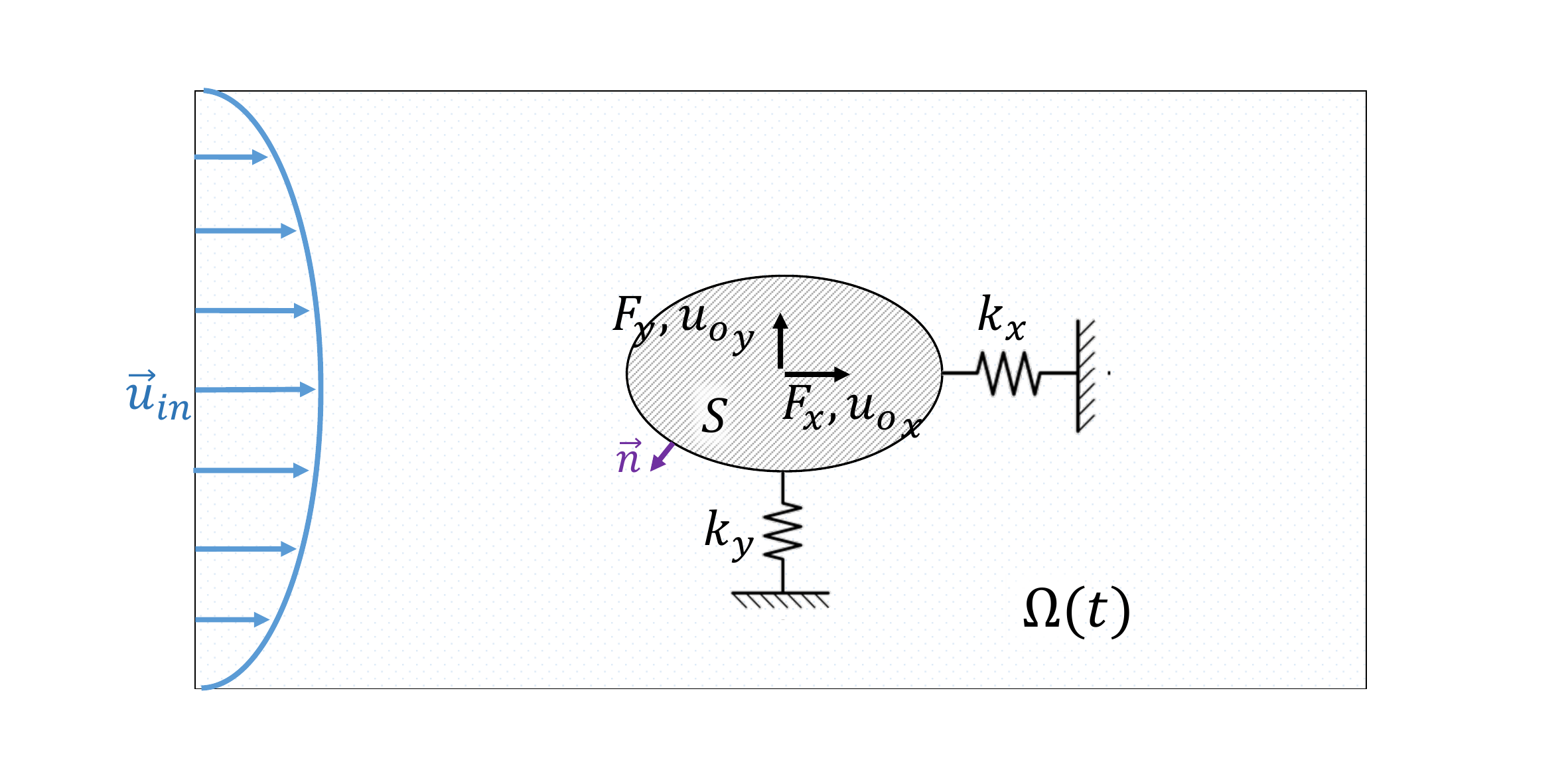}}
    
    \caption{Problem schematic representation: A non-deformable body with translational degrees of freedom along $x,y$, subject to an incompressible channel flow.}
 \label{fig:schematic}
  \end{center}
\end{figure}

As schematically depicted in \Cref{fig:schematic}, the problem consists of an incompressible fluid flow over a non-deformable body that can oscillate along the streamwise and transverse ($x$ and $y$) directions. The equations of motion for the solid ($s$) displacement $\mathbf{d}_s$ are

\begin{equation}
\label{osc}
  \rho_s A_s \partial_{tt} \mathbf{d}_s + K\mathbf{d}_s= (\rho_s-\rho_f) A_s\vec g -
  \int_{\partial\cal S} \mathbf{\sigma}(\mathbf{u},p)\vec n\,\text{d}s,
\end{equation}
\noindent where $\vec n$ is the unit normal vector on $\partial\cal S$ pointing from the solid to the fluid and $K$ is a diagonal matrix with the spring constants $k_x,\; k_y$. The cross-sectional area of the solid is $A_s$, the solid and fluid densities are $\rho_s$ and $\rho_f$ respectively and $\sigma$ is the 2-dimensional fluid stress tensor. 

The integral term of \eqref{osc} encodes the dynamic coupling condition between the fluid and solid subsystems, such that the equations of motion are essentially two uncoupled, externally forced oscillations. Internal damping has been set to zero to encourage the onset of vortex-induced vibrations.

For the fluid flow, we focus on the regime of 2D incompressible, laminar flows

\begin{equation}
\label{NS1}
  \nabla \cdot\mathbf{u}=0,\quad \rho_f (\partial_t \mathbf{u} + (\mathbf{u}\cdot\nabla)\mathbf{u})
  = \nabla \cdot\mathbf{\sigma} + \rho_f \vec g,
\end{equation}
\noindent with velocity $\mathbf{u}$, pressure $p$, kinematic viscosity $\nu_f$, gravity acceleration $\vec g$ and density $\rho_f$. The stress tensor $\sigma$ for a Newtonian fluid is:
\begin{equation}
\label{sigma}
  \mathbf{\sigma} = \rho_f\nu_f(\nabla\mathbf{u}+\nabla\mathbf{u}^T)-pI,
\end{equation}

\noindent where $I$ is an appropriate identity operator. The imposed boundary conditions to the flow are the kinematic coupling condition at the fluid/solid interface, the velocity at the inlet and the ``do-nothing" condition at the outlet \cite{Richter2017}

\begin{equation}
\label{BCs}
\left\{\begin{array}{ll}\mathbf{u}=\partial_{t} \mathbf{d}_s \; on \; \Omega(t) \cap S\\
\mathbf{u}=\mathbf{u}_{in} \; at \; x=0,\\
\rho_f \nu_f \nabla \mathbf{u} \mathbf{n}-pI=0 \; at \; x=L
\end{array}\right.
\end{equation}

From the Eulerian perspective, the fluid domain $\Omega(t)$ is changing over time (see \Cref{fig:schematic}). Through the ALE framework \cite{Dowell2001}, the Navier Stokes are mapped to a reference domain $\hat{\Omega}$, which could be selected to be the domain configuration at $t=0$. The map from the reference domain to the current configuration is then: $\hat{\mathbf{x}} \mapsto \mathbf{x}=\hat{\mathbf{x}}+\hat{\mathbf{d}}(t)$. The deformation field of the current configuration with respect to the reference one is given through function $\hat{\mathbf{d}}(t)$ (ALE map). For an analytical discussion of possible definitions of the ALE map, the reader is directed to Chapter 2.5 of \cite{Richter2017}. Considering modest gird deformations, the ALE map $\hat{\mathbf{d}}(t)$ is computed through the solution of a Laplace problem

\begin{equation} 
\label{laplace}
-\operatorname{div}\left(\hat{\nabla} \hat{\mathbf{d}}\right)=0 \quad \text { in } \hat{\Omega} \times(0, T],
\end{equation}

\noindent with boundary conditions on the interface $\hat{\Omega} \cap S$, and on the domain external boundaries:

\begin{equation}
\label{lapBC}
\left\{\begin{array}{ll}\hat{\mathbf{d}}(t)=\mathbf{d}_s(t)-\mathbf{d}_s(0) \; \text{on} \; \hat{\Omega} \cap S\\
\hat{\mathbf{d}}(t)=\mathbf{0} \; \text{on domain bounds}.
\end{array}\right.
\end{equation}

We notice that using definition \eqref{laplace}, the only time dependence of the ALE map originates from the boundary conditions \eqref{lapBC} and that $\hat{\mathbf{d}}$ depends solely on the solid deformation $\mathbf{d}_s(t)$. This observation will be proven useful in the ROM level, since the ALE map can be inverted at a very low added computational cost. Denoting $F$ as the Jacobian of the ALE map, and $J=\det{F}$, we get the ALE formulation for the Navier-Stokes equations \cite{Richter2017,Nonino2021}:

\begin{multline} 
\label{NS_ALE}
\left\{\begin{array}{l}
\rho_{f} J\left(\partial_{t} \hat{\mathbf{u}}+\hat{\nabla} \hat{\mathbf{u}} F^{-1}\left(\hat{\mathbf{u}}-\partial_{t} \hat{\mathbf{d}}\right)\right)-\\
-\operatorname{div}\left(J \hat{\sigma}\left(\hat{\mathbf{u}}, \hat{p}\right) F^{-T}\right)=J \rho_f \vec g \quad
\; \\
\; \; \; \; \; \; \; \; \; \; \; \; \; \; \operatorname{div}\left(J F^{-1} \hat{\mathbf{u}}\right)=0 \quad
\end{array}\right.
\end{multline}

From the model reduction point of view, it is of particular interest to construct the fluid dynamics model on domain $\hat{\Omega}$, following a different approach from \cite{Liberge2010}. In this way, we would isolate the flowfield snapshots from the solid motion, thus avoiding typical issues of transport-dominated equations projection \cite{Neeraj} close to the FSI interface. The ALE formulation \eqref{NS_ALE} of the Navier-Stokes is suitable for this task. However, the corresponding structure of the equations is no longer quadratic, as in typical, fixed domain Navier-Stokes \cite{Benner2021}. Instead, the structure of \eqref{NS_ALE} is highly-nonlinear. 

If we assume that the grid deformation is small, we can approximate $F \approx I$ and $J \approx 1$. This simplifying assumption is made to reveal an approximate model structure on the data-driven level, however it could constrain the data-driven model towards modest grid deformations. Such a limitation is either way dictated by the usage of the ALE map \eqref{laplace}. Simplifying \eqref{NS_ALE} based on the above, we get

\begin{multline} 
\label{NS_ALE1}
\left\{\begin{array}{l}
\rho_{f} \left(\partial_{t} \hat{\mathbf{u}}+\hat{\nabla} \hat{\mathbf{u}} \left(\hat{\mathbf{u}}-\partial_{t} \hat{\mathbf{d}}\right)\right)- \\
-\operatorname{div}\left( \hat{\sigma}\left(\hat{\mathbf{u}}, \hat{p}\right) \right)= \rho_f \vec g \quad \text { in } \hat{\Omega} \times(0, T]\\
\; \; \: \; \; \; \; \; \; \; \; \; \; \operatorname{div}\left( \hat{\mathbf{u}}\right)=0 \quad \text { in } \hat{\Omega} \times(0, T] \\
\end{array}\right.,
\end{multline}

\noindent while the pressure can be computed by taking the divergence of \eqref{NS_ALE1}:

\begin{multline} 
\label{Poisson}
-\frac{1}{\rho_f} \Delta p =\frac{\partial u}{\partial x}\frac{\partial}{\partial x}\left( u-\partial_{t}\hat{d}_x\right)+\frac{\partial u}{\partial y}\frac{\partial}{\partial x}\left(v-\partial_{t} \hat{d}_y\right)+\\ 
+\frac{\partial v}{\partial x}\frac{\partial}{\partial y}\left( u-\partial_{t} \hat{d}_x\right)+\frac{\partial v}{\partial y}\frac{\partial}{\partial y}\left( v-\partial_{t} \hat{d}_y\right)
\end{multline}

\noindent The structure of \eqref{NS_ALE1}, as well as the algebraic link between velocity and pressure \eqref{Poisson} will comprise the foundation towards the fluid data-driven model.

%%%%%%%%%%%%%%%%%%%%%%%%%%%%%%%%%%%%%%%%%%%%%%%%%%%%%%%%%%%%%%%%%%%%%%%%%%%%%%%%

\section{Methodology}
\label{sec:meth}

\subsection{Data-Driven Fluid Model Structure}%
\label{subsec:struct}

Discretizing equation \eqref{NS_ALE1} of the previous section with given time-implicit and spatial schemes, results in a quad\-ratic-bilinear structure for the evolution of the velocity flowfield $\mathbf{u}$ from timestep $k$ to timestep $k+1$

\begin{multline} 
\label{quadbil1}
\hat{\mathbf{u}}^{k+1}=A_u \hat{\mathbf{u}}^k + H_u \hat{\mathbf{u}}^{k+1} \otimes \hat{\mathbf{u}}^{k+1} + K_u \; \partial_{t} \mathbf{d}_s \otimes \hat{\mathbf{u}}^{k+1} + \\
+ Q_u \mathbf{p}^k + B_u \; \partial_{t} \mathbf{d}_s + L_u \; \hat{\mathbf{u}}_{in},
\end{multline}

\noindent where $A_u$ encodes information from the viscous term and the time derivative discretization in \eqref{NS_ALE1}. Through the Kronecker product, quadratic ($H_u$) and bilinear ($K_u$) terms are included, corresponding to the advection term of \eqref{NS_ALE1}. $Q_u$ encodes pressure dependence, while terms $B_u$, $L_u$ are introduced to enforce the kinematic boundary conditions on the solid-fluid interface and the inlet. From here on, we refer to the velocity flowfield on the reference configuration and thus notation $\; \hat{} \;$ is dropped in the following. All involved operators are sparse, originating from the discretization of the operators in \eqref{NS_ALE1}. 
The model is written in an implicit way; apart from the linear terms, all right-hand-side velocity terms are taken at the $\left(k+1\right)$st timestep. This is related to the coupling of fluid and solid subsystems, later discussed in this work. 

Furthermore, discretizing \eqref{Poisson} in a similar way reveals the algebraic relation of pressure with velocity, extensively discussed in \cite{Benner2021}

\begin{equation}
\label{pdisc}
E_p \mathbf{p}^k= A_p \mathbf{u}^k + H_p \mathbf{u}^k \otimes \mathbf{u}^k + K_p \; \partial_{t} \mathbf{d}_s \otimes \mathbf{u}^k + C_p,
\end{equation}

\noindent where $A_p$ originates from the outflow ``do nothing" boundary condition and $C_p$ is the reference pressure. Matrices $E_p, A_p, H_p, K_p$ are banded and sparse, based on the employed spatial discretization stencil.

The inverse of $E_p$ can be shown to be band-dominated \cite{Bickel2012}, while the corresponding band can be reduced through several re-ordering methods, e.g. \cite{Reid2006}.  We thus assume that after substituting equation \eqref{pdisc} in \eqref{quadbil1}, the resulting operators can be approximated by sparse matrices, with non-zero elements only in the positions corresponding to adjacent nodes of any given internal grid node and thus, any matrix row.

Finally, the structure of the data-driven model for the flowfield velocities on a reference domain takes the following, quadratic-bilinear form

\begin{multline} 
\label{quadbil2}
\mathbf{u}^{k+1}=A \mathbf{u}^k + H \mathbf{u}^{k+1} \otimes \mathbf{u}^{k+1} + K \; \partial_{t} \mathbf{d}_s \otimes \mathbf{u}^{k+1} + \\
 + B \; \partial_{t} \mathbf{d}_s + L \; \mathbf{u}_{in} + C,
\end{multline}

\noindent with sparse operators $A$, $H$, $K$ and a priori known operators $L$, $B$ for the enforcement of the Dirichlet boundary conditions at the flow inlet and fluid-solid interface.

\subsection{Average Flowfield Removal}%
\label{subsec:avg}

\begin{figure}[!htbp]
  \begin{center}
    \resizebox{8cm}{!}{
    \includegraphics{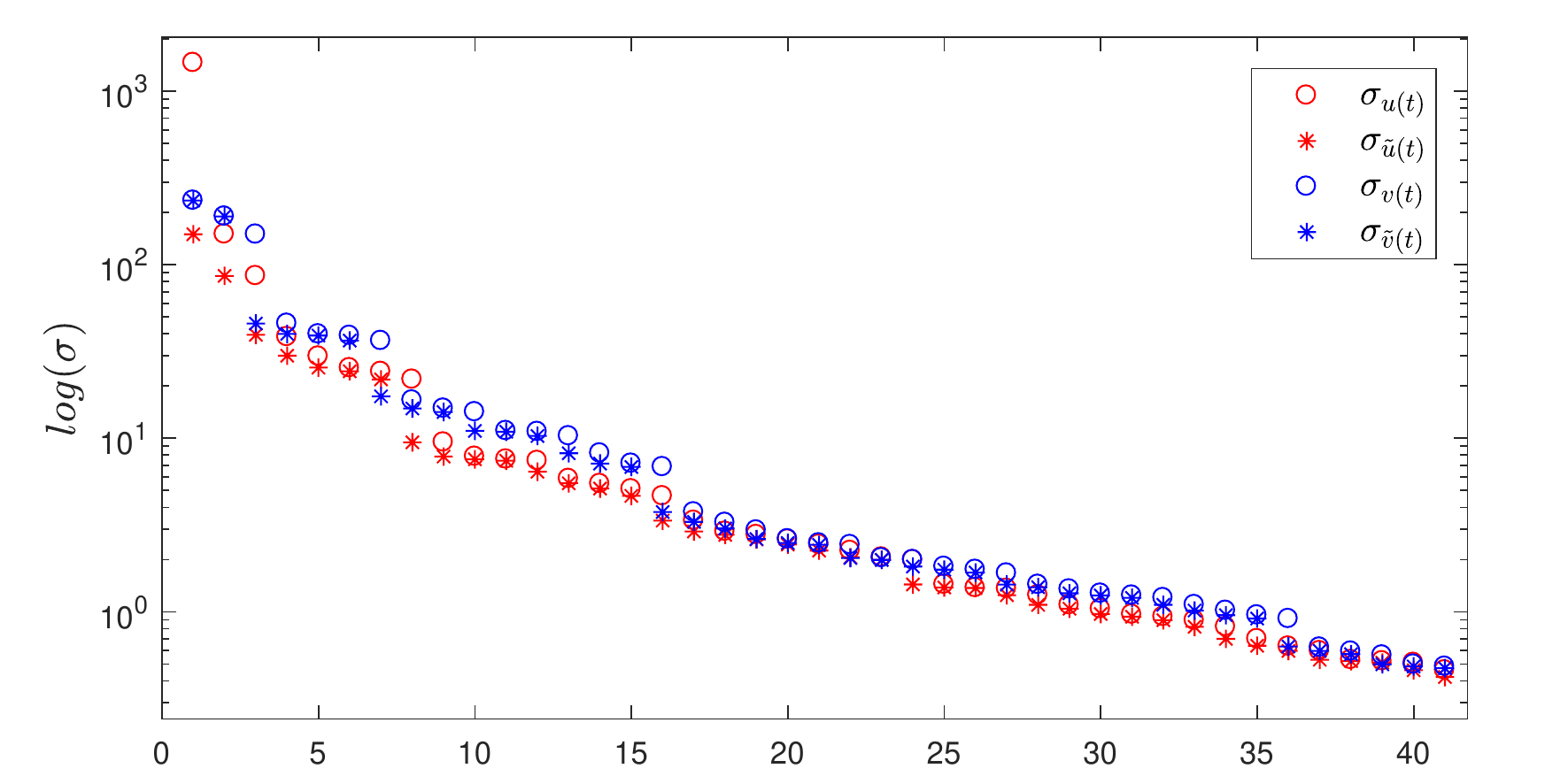}}
    
    \caption{Singular value decomposition of 3000 $\times$ 500 data matrices of the $u_x$, $u_y$ velocity components of a 2D flowfield for a VIV case. The first singular value of $u_x$ corresponds to the mean flow.}
 \label{fig:svd}
  \end{center}
\end{figure}

We hereby make a note on the decomposition of the flow into a dynamical and a stationary, average flowfield. For the flow around an oscillating cylinder, the mean flowfield $\bar{\mathbf{u}}$ over the reference domain $\hat{\Omega}$ is omnipresent and includes a significant part of the flow energy. This can be viewed in \Cref{fig:svd}, where the singular values of a 3000 $\times$ 500 flowfield snapshot matrix is plotted for velocity components $u_x$ and $u_y$, before and after removing the corresponding average of the time series. It is observed that the first $u_x$ singular value corresponds to the average $\bar{u}_x$, indicated by the observed singular values shift for $\tilde{u}_x(t)=u_x(t)-\bar{u}_x$. This effect is less evident for the $u_y$ component, since the average $\bar{u}_y$ is almost zero in this case.

\noindent As a result, we write the velocity field as:

\begin{equation}
\label{udec}
\mathbf{u}(t)=\bar{\mathbf{u}}+\tilde{\mathbf{u}}(t).
\end{equation}

Excluding the mean $\bar{\mathbf{u}}$ is known to have a beneficial effect \cite{Mendez2017} on the numerical manipulation of the data. As indicated by \Cref{fig:svd}, removing the mean reduces the condition number of the corresponding data matrix. Thus, assuming we have state access over some training time $[0,T_1]$, we subtract the time-average velocity field over this interval from the data and build a ROM only for the $\tilde{\mathbf{u}}(t)$ component. From this point and on, we refer to the prediction of $\tilde{\mathbf{u}}(t)$, dropping the $\; \tilde{} \;$ notation for simplicity. It can be observed that subtracting $\bar{\mathbf{u}}$ in \eqref{quadbil2} by substituting \eqref{udec} leaves the prescribed sparse, quadratic-bilinear structure unchanged.

\subsection{Grid Deformation}
\label{griddef}

Model \eqref{quadbil2} encodes the physics-based structure of the moderately-deformed domain, incompressible Navier-Stokes ALE formulation in \eqref{NS_ALE1}. Thus, it corresponds to the flow solution on a reference configuration. Before examining the data-driven solution of \eqref{quadbil2} for the reference configuration, we need to complement the model with a map from the reference domain to the current, deformed one. 

For the non-intrusive modeling procedure, the flow velocity data over some training time $t=[0,T_1]$ is imported, on a given grid $\Omega (t)$. The grid adjacency information is needed for the application of the aforementioned adjacency-based sparsity pattern.

The deformation of the grid over $[0,T_1]$ is often also provided in case of proprietary software. However, different solvers employ variations of \eqref{laplace} to derive the grid deformation \cite{Nonino2021, Richter2017}, often not traceable by the end-user.  Also, in the case of experimental data (e.g. PIV campaigns \cite{Mendez2017}) there is no notion of grid deformation. 

Thus, it is necessary to construct a new fluid mesh and compute its deformation based on a predefined ALE map (in our case \eqref{laplace}). Hence, the imported velocity data need to be interpolated at each training timestep from the solver grid positions (or e.g. PIV domain in experimental campaigns), to the positions of a re-computed grid, based on the selected ALE map. For the constructed mesh, grid adjacency information is stored, allowing for the application of an adjacency-based sparsity pattern in the non-intrusive model.

The motion of each grid node is given by discretizing and solving \eqref{laplace}. As in standard practice, we denote $\Lambda$ as the discretized Laplace operator, augmented by identity matrix rows for the Dirichlet boundary conditions in \eqref{lapBC}. $I$ denotes the set of nodes on the interface $\Omega(t) \cap S$, which should follow the motion of the solid (see eq. \eqref{lapBC}). The displacement of the grid nodes is then given by

\begin{equation} 
\label{disc_Lap}
\hat{\mathbf{d}}=\left(\sum_{i \in I}{\Lambda^{-1}_{:,i}}\right)(\mathbf{d}_s(t)-\mathbf{d}_s(0))^T,
\end{equation}

\noindent where the reference configuration is taken as the one at $t=0$. We observe that $\left(\sum_{i \in I}{\Lambda^{-1}_{:,i}}\right)$ should be computed once and then the grid deformation can be computed directly for a given time $t$, based on the solid motion ($(\mathbf{d}_s(t)$) prediction. In essence, \eqref{disc_Lap} gives an analytical expression of a basis for the grid displacement $\hat{\mathbf{d}}$ with respect to the solid displacement.

\subsection{Data-driven, Adjacency-based Sparsity}
\label{adj}

The task in the case of non-intrusive modeling is to infer the unknown operators of \eqref{quadbil2}, i.e. to construct a data-driven model for the 2D velocity field. Typically, the model operators are inferred by first projecting the data to the leading SVD modes of the data matrix \cite{Schmid2011, Peherstorfer2016}. However, we hereby follow a different approach, motivated by the discussed adjacency-based sparsity of the discretized operators which we want to infer. 

Following this approach is also accommodated by the application-based need for grid generation, already discussed in \Cref{griddef}. In particular, using the grid adjacency information we can identify adjacent nodes and extract corresponding velocity data. By examining each degree of freedom, we define a sparsity pattern for the full-order, inferred operators. This idea, briefly analyzed in \cite{pidmd}, is here applied to all internal nodes of the 2D grid, for the prediction of $u_x$ and $u_y$ velocity components.

Enforcing sparsity enables computing the full-order, data-driven operators in \eqref{quadbil2}, since only neighbouring node products of terms $K \; \partial_{t} \mathbf{d}_s \otimes \mathbf{u}^{k+1}$ and $H \mathbf{u}^k \otimes \mathbf{u}^k$ are non-zero. This also allows enforcing uniqueness of the data-driven model with respect to the states (velocities), by avoiding commutative velocity products. Such a property cannot be enforced when constructing the non-intrusive model in the reduced space, since the reduced vector elements consist of linear combinations of the states. None\-theless, it is evident that the offline cost of this approach scales with the number of nodes on the interpolated grid and thus is expected to be significantly elevated, compared to that of projection-first methods.

In practice, to apply the aforementioned sparsity pattern based on grid adjacency we examine one row of \eqref{quadbil2} corresponding to some internal node. Dirichlet boundary conditions for nodes at $x=0$ ($i_{in}$) and nodes on $F \cap S$ ($i_{s}$) can be a priori satisfied by setting $A_{i_{in},.}=H_{i_{in},.}=K_{i_{in},.}=0, \; C_{i_{in}}=0$ and $A_{i_{s},.}=H_{i_{s},.}=K_{i_{s},.}=0, \; C_{i_{s}}=0$, while entries of $1$ are registered in respective positions of $B, L$. Thus, matrices $B, L$ are grid-dependent, a priori known, permutation matrices.

We split the quadratic terms into two contributions $h_A$ and $h_B$; the first one includes velocity products including the velocity of the examined node, while the second one includes products between adjacent node velocities. Based on this, we now look more closely into a row of \eqref{quadbil2}. The equation for the $u_x$ or $u_y$ velocity of a node $i$ with a set of adjacent nodes $q(i)$ is then

\begin{multline} 
\label{qb_node}
u^{k+1}_i=\sum_{q(i)}{\big(a_{i,j}u^{k}_{j}+ \left[ h_{A_{i,j}} \; h_{A_{i+zn,j}} \right]  \left[ u^{k+1}_{i} u^{k+1}_{i+zn}\right]^T u^{k+1}_{j}}\\{+ \left[ k_{i,j} \; k_{i,j+2n} \right] u^{k+1}_{j} \partial_{t} \mathbf{d}_s + \sum_{q(i)}{\left( h_{B_{j,l}} u^{k+1}_{l} u^{k+1}_{j}\right)} \big) },
\end{multline}

\noindent where we set $z=1$ for $u_x$ and $z=-1$ for $u_y$ respectively. In this general form, the terms of $h_A, h_B$ are not unique. In practice, we eliminate the terms of $h_B$ involving the velocity components of node $i$ so that $h_A$ and $h_B$ share no common terms. We also eliminate non-unique, commutative terms in $h_B$. As a result, if node $i$ has $m$ neighbouring DOFs, then $h_{B_i}$ includes $m \choose 2$ terms \cite{Peherstorfer2016}.

Equation \eqref{qb_node} for node $i$ can be written for every timestep in $[0,T_1]$. This leads to the formulation of a least squares problem for the entries of operators $A, H, K$

\begin{equation} 
\label{LS_noreg}
\min _{a_{i,.}, h_{A_{i,.}},h_{B_{i,.}},k_{i,.}}\left\|\left[a_{i,.}, h_{A_{i,.}},h_{B_{i,.}},k_{i,.}\right]^T \mathcal{D}-\mathbf{u_i}^{k+1}\right\|_{2},
\end{equation}

\noindent where, following the notation of \cite{Benner2021},
\begin{equation}
\label{Dmat}
\mathcal{D}^T=
\begin{bmatrix}
  \mathbf{u_{q(i)}} \\
  \mathbf{u}_i \mathbf{u}_{q(i)} \\
  \mathbf{u}_{i+nz}\mathbf{u}_{q(i)} \\  
  \mathbf{u}_{q(i)}^2\\
  \mathbf{u}_{q(i)} \otimes \mathbf{d}_s\\
\end{bmatrix}
.
\end{equation}

The solution of \eqref{LS_noreg} for an internal DOF $i$ is a vector with the entries of line $i$ of the operators $A,H,K$. Based on grid adjacency information, these entries can then be stored in corresponding positions of the operators, thus leading to an adjacency-based sparsity pattern of the full-order, non-intrusive model.

\subsection{Regularization}
\label{Reg}

Adding regularization is required for the practical solution of \eqref{LS_noreg}, to avoid numerical errors due to small singular values of $\mathcal{D}$. We employ Tikhonov regularization \cite{Swischuk2020,McQuarrie2021} to penalize solutions with high $L_2$ norm by modifying each least squares problem as follows:

\begin{multline}
\label{LS_reg}
\min_{a_i, h_{A_i},h_{B_i},k_i}{\left({\left\| \left[a_i, h_{A_i},h_{B_i},k_i\right]^T \mathcal{D}-\mathbf{u_i}^{k+1}\right\|}_2+ \right.}     \\
        \left.+ {\lambda_1{\left\| \left[a_i,k_i\right] \right\|}_2+ \lambda_2 {\left\| \left[h_{A_i},h_{B_i}\right] \right\|}_2}\right).
\end{multline}

\begin{figure}[!htbp] 
  \begin{center}
    \resizebox{8.8cm}{!}{
    \includegraphics{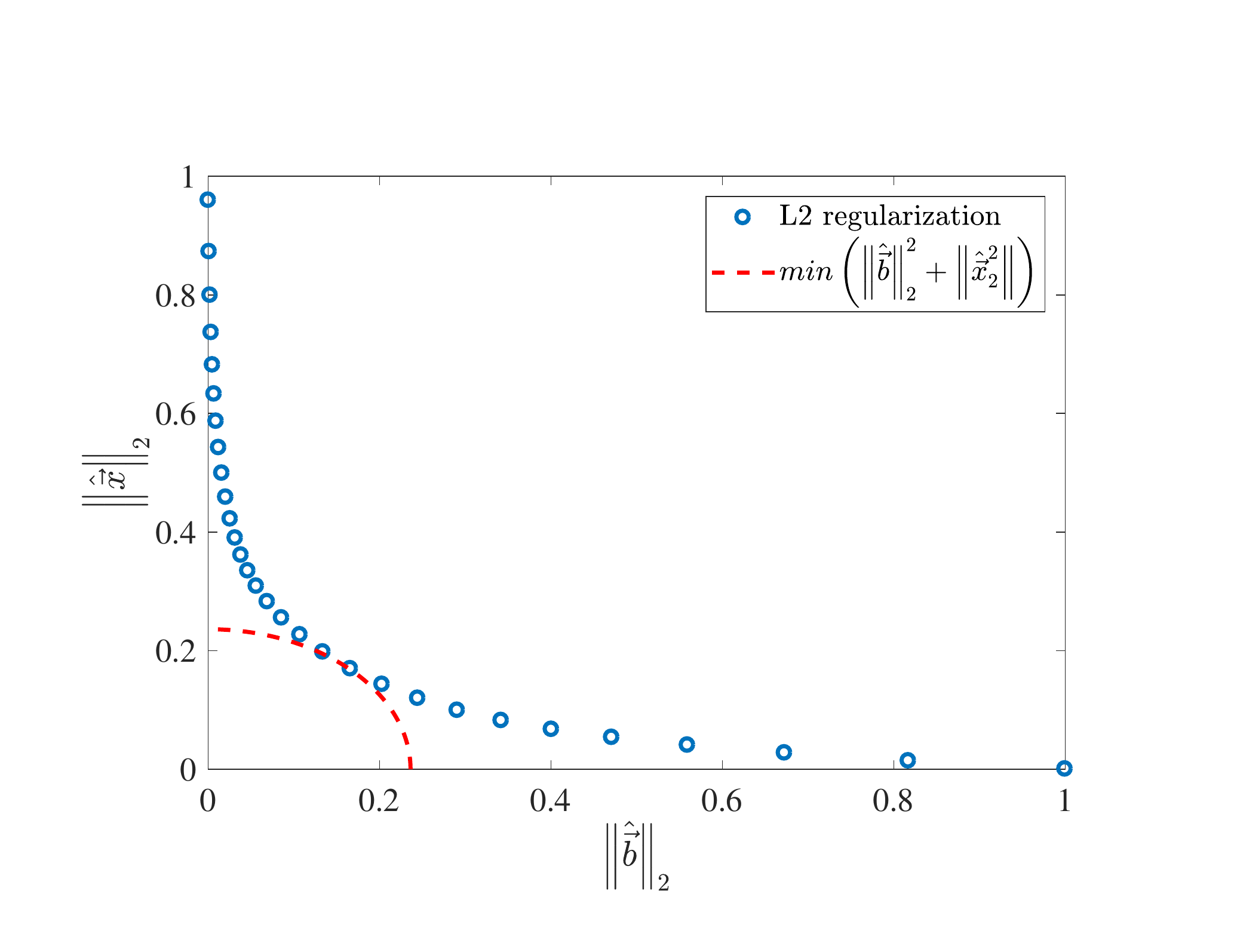}}
    \caption{Example of $L_2$ regularization L-curve criterion: The model (and thus the $\lambda$ value) which gives the smallest normalized distance from the $\left({\left\| \hat{\vec{b}} \right\|}_{2},{\left\| \hat{\vec{x}} \right\|}_{2}^2\right)$ origin is chosen.}
  \label{fig:L2_er}
  \end{center}
\end{figure}

For the current application, setting $\lambda_1=\lambda_2$ yields satisfactory results. Hence, for each degree of freedom, we solve problem \eqref{LS_reg}, for different values of $\lambda=\lambda_1=\lambda_2$. The optimal regularization parameter is chosen based on the L-curve criterion \cite{Hansen2000}. In particular, we aim to leverage between the least squares solution error and the solution norm. Denoting $\left\| \vec{b} \right\|_{2}$ as the least squares solution error and $\left\| \vec{x} \right\|_{2}$ as the solution norm, we seek for

\begin{equation} 
\label{criterion}
\min _{\lambda}{\left({\left\| \hat{\vec{b}} \right\|}_{2}^2+{\left\| \hat{\vec{x}} \right\|}_{2}^2\right)},
\end{equation}

\noindent where $\; \hat{} \;$ indicates normalization of the $\left\| \vec{b} \right\|_{2}$ and $\left\| \vec{x} \right\|_{2}$ values to $[0,1]$. \Cref{fig:L2_er} illustrates the procedure of finding the optimal regularization parameter for one grid node (e.g. for the $u_x$-velocity component): For each $\lambda$ value, we obtain a pair $(\vec{b},\vec{x})$. The $\lambda$ value for which the corresponding pair (denoted $(\vec{b}_{L},\vec{x}_{L})$) satisfies \eqref{criterion} is then selected.

In the full-order setting, we can reduce the computational cost by spatially interpolating the optimal regularization parameter. In essence, since velocity varies smoothly over the domain (dealing with an incompressible flow), matrix $\mathcal{D}$ in \eqref{Dmat} and thus its singular values are also expected to vary smoothly over the domain. We can thus compute the optimal regularization parameter at randomly sampled nodes and approximate $\lambda$ at the remaining nodes by interpolation.

By summing the contribution of $\left\| \vec{x} \right\|_{2}$ and $\left\| \vec{b} \right\|_{2}$ for each $\lambda$ value over all the internal DOFs, we can plot a global L-curve. The corresponding sums can be calculated for the optimal $\lambda$ (at each DOF), i.e. the training error of the model and norm of the inferred operators. The inferred model training error and operators norm will not necessarily lie on the global L-curve, since the optimal $\lambda$ is independently computed for each DOF.

\begin{figure}[!htbp] 
\centering
\subfloat[L-curve for the DOF-wise and interpolated $\lambda$ models: Slightly increased error for the interpolated $\lambda$ model, at a significantly lower cost.]{%
  \includegraphics[clip,width=0.9\columnwidth]{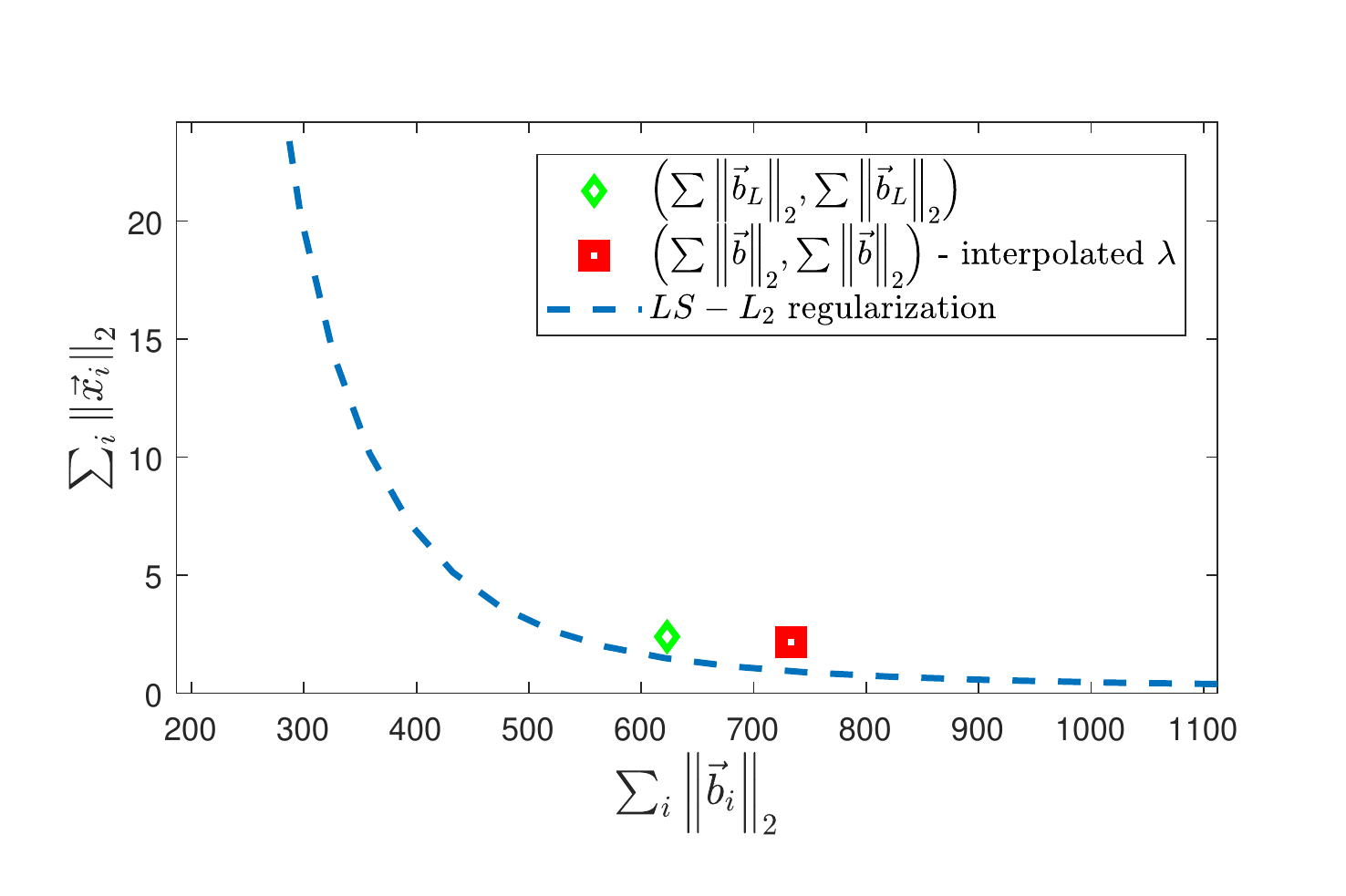}%
}

\subfloat[Optimal $\lambda$ computed at each grid node ($u_x$ velocity).]{%
  \includegraphics[clip,width=1.1\columnwidth]{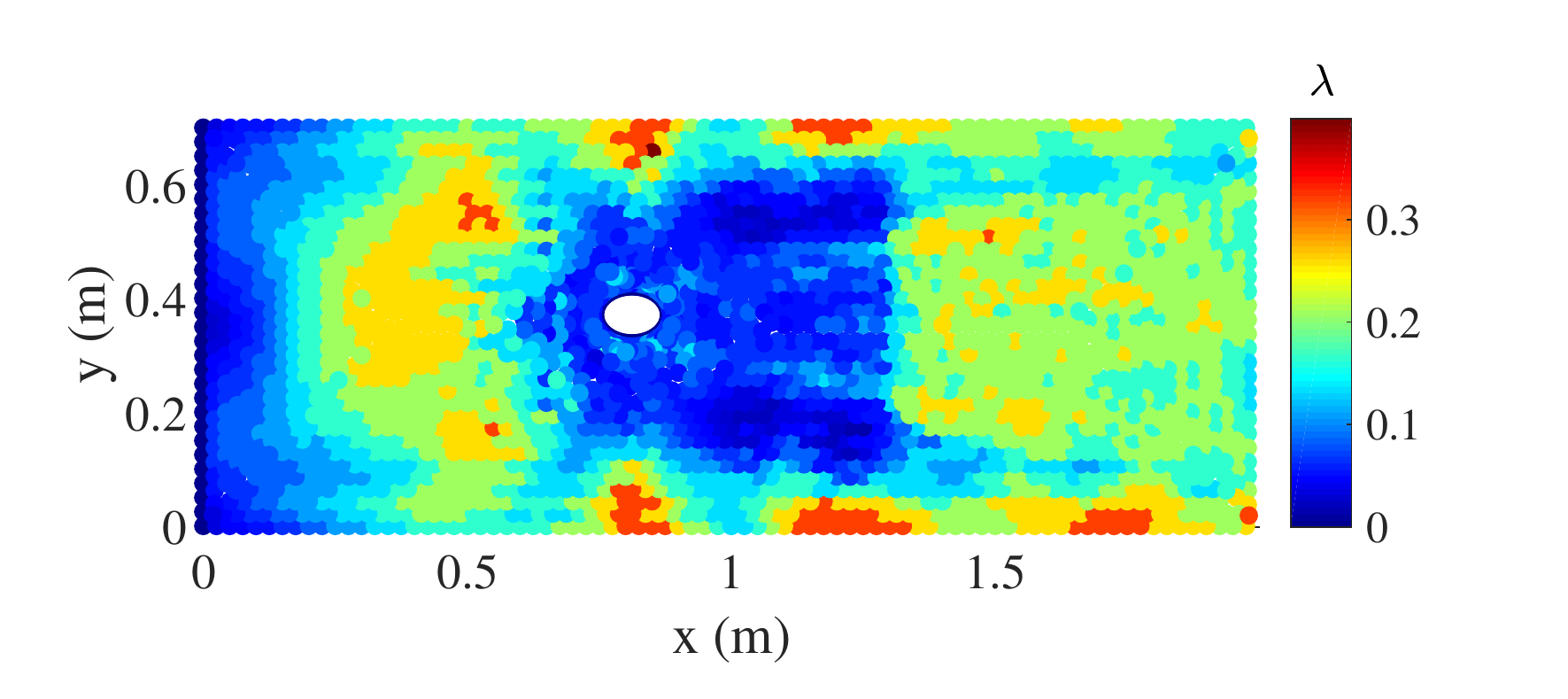}%
}

\subfloat[Optimal $\lambda$ computed at $10\%$ of grid nodes, interpolated for the rest $90\%$.]{%
  \includegraphics[clip,width=1.1\columnwidth]{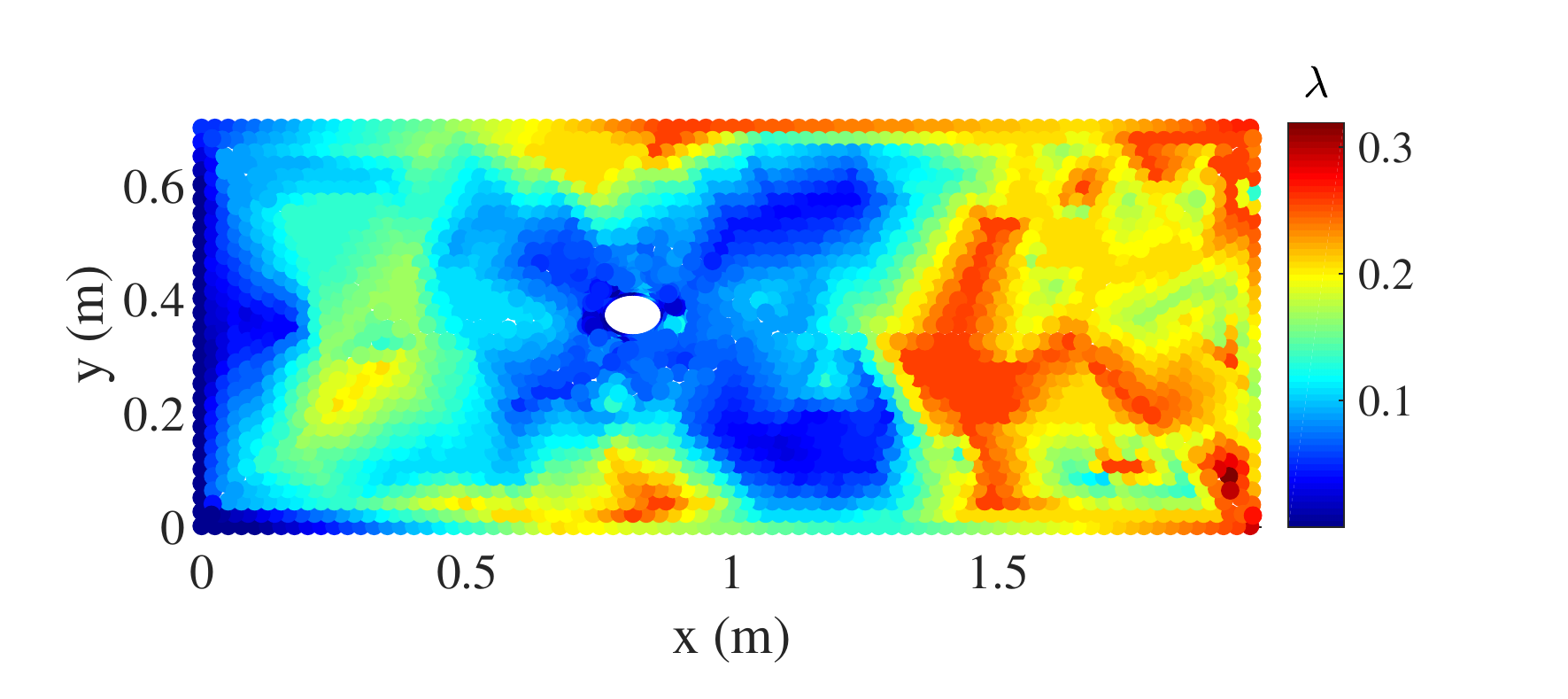}%
}

\caption{Regularization parameter selection for fluid nodes: Indicatory example for a $Re=180$ dataset (discussed in detail in \Cref{sec:res}). Increased $\lambda$ values in the flow wake for both the exact and the interpolated optimal regularization strategies.}
\label{fig:L2_field}
\end{figure}

\Cref{fig:L2_field} indicates the trade-off between the non-intrusive model training error and inferred operators norm, for a $Re=180$ flowfield, further analyzed in \Cref{sec:res}. By spatial interpolation of the regularization parameter, the least squares computational cost is significantly reduced. For example, if 20 $\lambda$ values are considered and the optimal $\lambda$ is computed for $10\%$ of the DOFs, then only one least squares problem should be solved for the remaining $90\%$ of the DOFs. This translates to a decrease in the involved least squares computational cost by a factor of $\approx 6.5$ (without considering the cost of $\lambda$ interpolation). A zoomed-in view of the global L-curve is depicted in \Cref{fig:L2_field}a. A minor increase in the model training error when interpolating $\lambda$ is recorded, compared to the model computed from solving \eqref{LS_reg} for all degrees of freedom. \Cref{fig:L2_field}b illustrates the optimal regularization parameter for each grid node based on \eqref{criterion}, for the $u_x$ velocity component. We observe that $\lambda$ is higher in the wake region of the flow, where the flow exhibits more complex dynamics. \Cref{fig:L2_field}c corresponds to the interpolated $\lambda$ case, computing the optimal parameter for $10\%$ of the internal grid nodes. \Cref{fig:L2_field}b and \Cref{fig:L2_field}c exhibit differences on the selected $\lambda$, however the resulting difference in training error is only minor. It is noted that parallelization of the least squares problems could also significantly reduce the offline clock time required for inferring the operators through the employed methodology, however, this has not been considered in this work.

At this step, the sparse, full-order, data-driven model of \eqref{quadbil2} has been computed. Next, the force induced from the flow to the solid should be predicted, as to couple the fluid and solid dynamics.
%%%%%%%%%%%%%%%%%%%%%%%%%%%%%%%%%%%%%%%%%%%%%%%%%%%%%%%%%%%%%%%%%%%%%%%%%%%%%%%%

\subsection{Surface Forces Modeling}
\label{SurfaceModel}

As shown in \eqref{osc}, the forcing from the fluid flow to the solid motion originates from the integral of stresses on the body surface (eq. \eqref{sigma}). Focusing on non-deformable solid motion along the two axes, a model for the forces $F_x, F_y$ (\Cref{fig:schematic}) is required. The mean velocity flowfield over the training time $[0,T_1]$ results in a constant force $\bar{\mathbf{F}}=\left[ \bar{F_x}, \bar{F_y} \right]^T$. As a result, we can similarly write the force as

\begin{equation}
\mathbf{F}=\bar{\mathbf{F}}+\tilde{\mathbf{F}}.
\end{equation}

By consideration of the examined application, $\bar{\mathbf{F}}$ should tend to zero as training time increases, since the solid motion is oscillatory. We aim to predict the dynamically evolving component $\tilde{\mathbf{F}}$, in line with the focus on the dynamically varying velocity flowfield component. In the following we drop $\; \tilde{} \;$ for simplicity.

We assume solid motion parameters (mass, spring constants) to be known. Mass and (linear) spring constants can be easily determined both numerically, but also experimentally through the impulse response for the solid oscillation. Therefore, only kinematic data is necessary for constructing the non-intrusive model. The velocity training data for the solid, as well as the known mass and spring constants are substituted in \eqref{osc} to obtain numerical training data for the integral forcing terms we aim to predict. 

 Based on the typical decomposition of flow-induced forces to a shear term (linear) and a pressure term (quadratic), we hereby employ a non-intrusive, quadratic model for the force, in the following form:

\begin{equation}
\label{force}
\mathbf{F}^{k}=A_F \mathbf{u}^k + H_F \mathbf{u}^k \otimes \mathbf{u}^k.
\end{equation}

\begin{figure}[!htbp] 
  \begin{center}
    \resizebox{8cm}{!}{
    \includegraphics{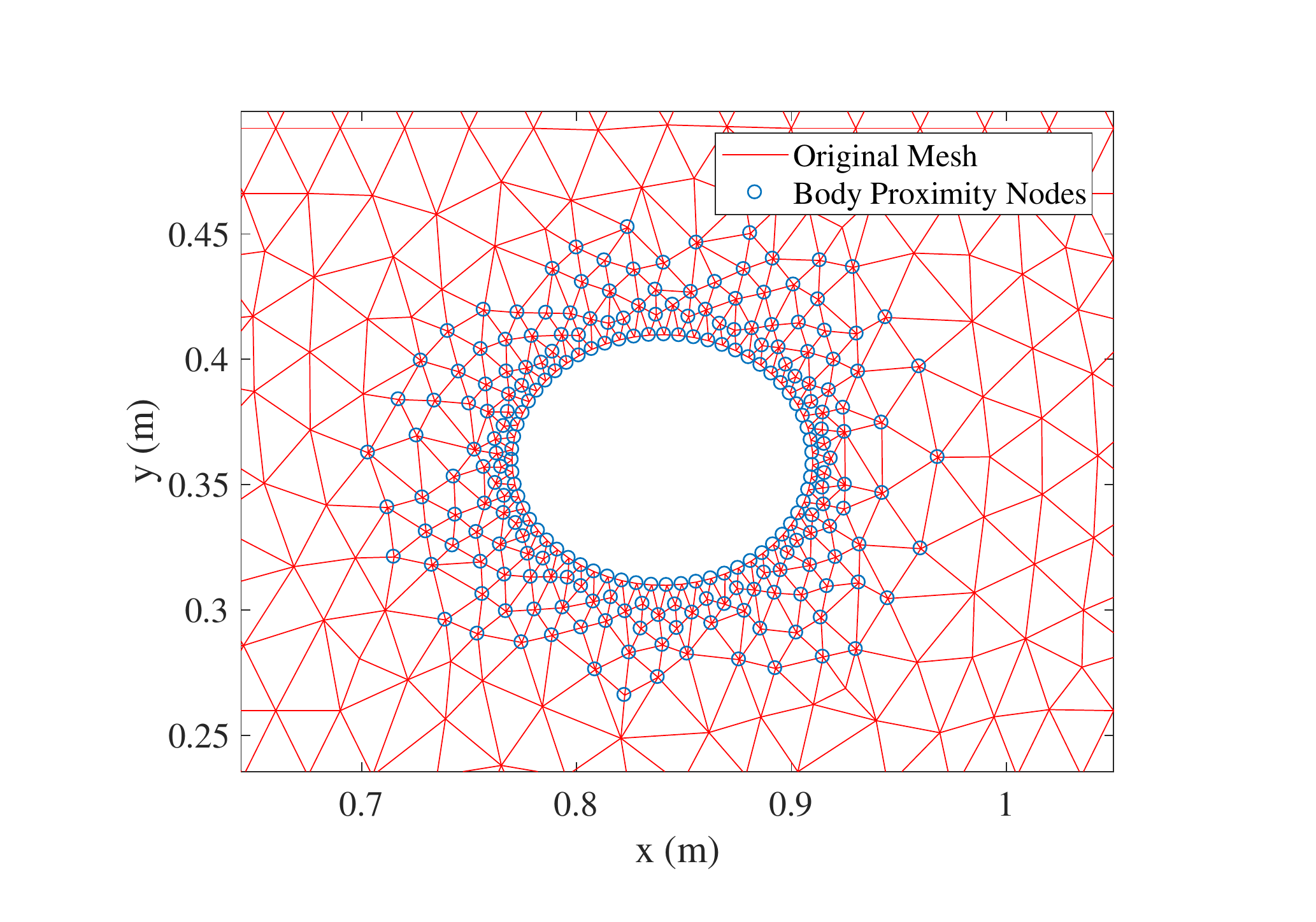}}
    \caption{Body-proximity nodes: Given a distance threshold, only $2n_F$ DOFs are included in the body forces non-intrusive model}
  \label{fig:F_grid}
  \end{center}
\end{figure}

For this part, a second adjacency matrix is formulated. Equations \eqref{osc} and \eqref{sigma} indicate that only the velocity of nodes sufficiently close to the body will contribute to the force acting on it. As a result, the full-order output model \eqref{force} has $2 n_F$ DOFs (velocities $u_x$ and $u_y$), where $n_F<<n$ is the number of nodes at the proximity of the interface $\hat{\Omega} \cap S$. An illustration of this proximity grid is given in \Cref{fig:F_grid}.

As before, not all $2n_F \choose 2$ terms of $\mathbf{u} \otimes \mathbf{u}$ were included in the model, but only the ones corresponding to adjacent pairs of DOFs. In fact, using only quadratic terms of the sort $u_i u_{q(i)}$, where $q(i)$ denotes the adjacent DOFs for each DOF $i$ and $u$ can be $u_x$ or $u_y$, proved to be sufficiently accurate for body forces prediction, while restraining the involved least square problems dimension.

As presented for the velocity flowfield in \eqref{adj} and \eqref{Reg}, we formulate the full-order matrices $A_F, H_F$ by solving an $L_2$-regularized least squares problem

\begin{equation}
\min _{A_F, H_F}{\left(\left\|\left[A_F \; H_F \right]^T \mathcal{D}_F - \mathbf{F} \right\|_{2}
+ \lambda_F\left\|\left[A_F,H_F\right] \right\|_{2}\right)},
\end{equation}

\noindent where $A_F,\; H_F$ are of dimension $2 \times 2n_F$.

\subsubsection{Coupled ROM-Oscillation Dynamics}
\label{proj}

Up to this point, the sparse, full-order, non-intrusive models for the velocity flowfield and the force on the solid have been computed. The non-intrusive ROM is then computed via projection, using a POD basis. 

The flowfield model is projected to the leading singular modes of data matrix $U$. The columns of $U$ are vectorized velocity data over the new fluid mesh (see \eqref{griddef}), stacked over training time $[0,T_1]$. Following the logic of Proper Orthogonal Decomposition, we compute the SVD $U=\Phi \Sigma \Psi ^T$ and truncate the leading $r$ singular values. Denoting $\Phi_r=\Phi_{.,1:r}$, we project the grid velocities $u_k$ to the leading $r$ modes, such that

\begin{equation} 
\label{phi}
u_k=\Phi_r \tilde{u}_k.
\end{equation}

Applying \eqref{phi} to model \eqref{quadbil2}, we obtain the reduced matrices as follows:

\begin{multline}
\tilde{A}=\Phi_r^T A \Phi_r, \; \tilde{H}=\Phi_r^T H \Phi_r \otimes \Phi_r, \; \tilde{K}=\Phi_r^T K I_{2x2} \otimes \Phi_r \\
\tilde{B}=\Phi_r^T B, \; \tilde{L}=\Phi_r^T L, \; \tilde{C}=\Phi_r^T C.
\end{multline}

We observe that due to the sparsity of matrix $H$, we can efficiently perform the projection to $\tilde{H}$. In particular, we split matrix $H$ into $H_A$ with entries including the ``self node" terms (see eq. \eqref{qb_node}) and $H_B$ containing unique products of neighbouring node velocities. We store only the non-trivial entries of these two matrices; as a result, we need to store matrices of dimensions $2n \; \times \; 2\max_{i}{q(i)}$ and $2n \; \times \; {2\left(\max_{i}{q(i)}-1\right) \choose 2}$, respectively for $H_A$ and $H_B$, where $n$ is the total number of nodes (velocities $u_x$ and $u_y$). This allows using a different computational stencil for each of the two terms $H_A$ and $H_B$. In this application, we use the second-degree adjacent nodes for $H_A$ and the first-degree adjacent nodes for $H_B$, limiting the number of required Kronecker products. Since the two matrices share no common non-zero entries, we can then write

\begin{equation} 
\label{quadproj}
H \Phi_r \otimes \Phi_r = H_A \Phi_r \otimes \Phi_r +  H_B \Phi_r \otimes \Phi_r.
\end{equation}

Knowing the sparsity pattern of $H_A$ and $H_B$, we compute only the relevant Kronecker products of $\Phi_r \otimes \Phi_r$ for each case. Once matrix $H \Phi_r \otimes \Phi_r$ is computed by the above procedure, it is multiplied on the left with $\Phi_r^T$. This procedure significantly reduces the number of operations required for the projection of the quadratic matrix.

For the force acting from the flow to the solid, we follow a similar procedure: We identify the lines of $\Phi_r$ corresponding to the nodes close to the body from the grid adjacency information (as shown in \Cref{fig:F_grid}). We denote $\Phi_s$ the $2n_F\times r$ matrix by retaining the necessary lines of $\Phi_r$. Then, the projection step for \eqref{force} is

\begin{equation}
\tilde{A}_F= A_F \Phi_s, \; \tilde{H}_F= H_F \Phi_s \otimes \Phi_s.
\end{equation}

Having obtained a non-intrusive model for both the velocity field and the resulting forces on the solid along the streamwise and transverse direction, we are ready to couple the fluid (data-driven) model with the solid (first-principle) one. The data-driven ROM writes as

\begin{multline}
\label{flowROM}
\left\{\begin{array}{l}
\tilde{\mathbf{u}}^{k}=\tilde{A} \tilde{\mathbf{u}}^{k-1} + \tilde{H} \tilde{\mathbf{u}}^k \otimes \tilde{\mathbf{u}}^k + \tilde{K} \partial_t \mathbf{d_s}^{k} \otimes \tilde{\mathbf{u}}^k + \\
\; \; \; \; \; \; \; \; \; +\tilde{B} \partial_t \mathbf{d_s}^{k} + \tilde{L} \mathbf{u}_{in} + \tilde{C} \\
\mathbf{F}^{k}=\tilde{A}_F \tilde{\mathbf{u}}^k + \tilde{H}_F \tilde{\mathbf{u}}^k \otimes \tilde{\mathbf{u}}^k,
\end{array}\right.
\end{multline}

\noindent while the solid oscillation equations \eqref{osc} are integrated using a Crank-Nicholson scheme, with timestep $\Delta t$:

\begin{multline}
\label{CrNic}
\left\{\begin{array}{l}
\partial_t \mathbf{d_s}^{k}+\frac{\Delta t K}{2 \rho_s A_s} \mathbf{d_s}^{k}-\frac{\Delta t}{2} \mathbf{F}^k\\
=\partial_t \mathbf{ d_s}^{k-1}-\frac{\Delta t K}{2 \rho_s A_s} \mathbf{d_s}^{k-1}+ \Delta t \; \mathbf{g}+\frac{\Delta t}{2} \mathbf{F}^{k-1} \\
\;\\
\mathbf{d_s}^{k}-\frac{\Delta t}{2} \partial_t \mathbf{ d_s}^{k} \\
=\mathbf{d_s}^{k-1}+\frac{\Delta t}{2} \partial_t \mathbf{d_s}^{k-1}
\end{array}\right.
\end{multline}

An interesting observation during simulation was that the known stiff coupling between the fluid and solid subsystems in FSI numerical simulation \cite{Richter2010, Cori2015} transfers also to this data-driven/first principle two-way coupling, rendering an explicit model unstable. Thus, the model is constructed with an implicit formulation, requiring convergence at each timestep. In detail, the force $\mathbf{F}^{k}$ resulting from the data-driven flowfield ROM $\tilde{\mathbf{u}}^{k}$ is driving the solid oscillation. The resulting solid displacement $\mathbf{d}_s^{k}$ is in turn affecting the surrounding flow $\tilde{\mathbf{u}}^{k}$. A combined convergence criterion of ${\left\| \tilde{\mathbf{u}}_{j+1}^k -\tilde{\mathbf{u}}_{j}^k\right\|}_2<10^{-12}$ and ${\left\| \mathbf{d_s}_{j+1}^k -\mathbf{d_s}_{j}^k\right\|}_2<10^{-12}$  turned out to provide good results, where $j$ denotes the index of iterations within timestep $k$. For each implicit iteration, a typical successive relaxation method was used \cite{Revstedt}, with $\tilde{\mathbf{u}}_{\ast}^k=(1-\omega) \tilde{\mathbf{u}}_{j}^k+\omega \tilde{\mathbf{u}}_{j+1}^k$. During simulation, a maximum of $4-5$ iterations per timestep where required to reach convergence.

\section{Simulation Results}%
\label{sec:res}

The above methodology was coded in MATLAB and applied to two numerical testcases for the VIV of an ellipse-shaped solid. The details for the two testcases are given in \Cref{tab1}. The solid has streamwise and transverse eigenfrequencies (in vacuum) of $f_s=4.38\;Hz$ and is subject to a laminar, incompressible channel flow. A parabolic velocity profile is prescribed at the channel inlet, with a maximum velocity of $u_{max}=1.5\bar{u}_{in}$, while the domain size is $4 \; \times \; 1\; m^2$. 

The reference length is twice the average of the semiaxes $D=R_x+R_y$ and the reference velocity $u_\infty$ is the average inlet velocity. The two testcases then correspond to $Re=90$ and $Re=180$. A low mass ratio is also selected ($\rho_s/\rho=1.2$), to promote a wide VIV lock-in velocity range and considerable peak amplitude \cite{Williamson2004}. The added mass is $m_A=0.011\;kg$, which leads to eigenfrequencies of the submerged solid of $f_N\approx3.24\; Hz$. The reduced velocity $U^\ast=u_\infty /(f_N D) $ in the two testcases is then $U^\ast=1.93$ and $U^\ast=3.86$. This indicates that they respectively belong to the right and left parts of the initial VIV branch \cite{Riches2018}. All numerical simulations were performed with the GASCOIGNE open-source software \cite{GASCOIGNE} on a computational grid of $10256$ nodes and simulation time of $15$ and $10$ seconds respectively for $Re=90,\;180$.

\begin{table}[!htbp]
  \caption{VIV simulation properties: Two testcases with $Re=90, 180$ were examined, at a low mass ratio $\rho_s/\rho=1.2$.  }
  \label{tab:example}
  
  \begin{center}
    \begin{tabular}{llllc}
      $\nu (m^2/s)$ & $\rho (kg/m^3)$ & $\rho_s (kg/m^3)$ & $R_x (m)$ & $R_y (m)$\\
      0.001 & 1 & 1.2 & 0.07 & 0.05 \\
      \hline\noalign{\medskip}
      $k_x (N/m)$ & $k_y (N/m)$ & $\bar{u}_{in}(m/s)$ & $\Delta t$\\
      10 & 10 & 0.75/1.5 & 0.05\\
      \noalign{\medskip}\hline\noalign{\smallskip}
    \end{tabular}
  \end{center}
  \label{tab1}
\end{table}

The grid interpolation step described in \eqref{griddef} was performed focusing on a region close to the body. An unstructured mesh of $2085$ nodes is constructed on a truncated domain of $[0.2,2.4] \times [0.1,0.9]$, with the use of the MESH2D MATLAB tool \cite{mesh2d}. The data from the last $5.9$ seconds of simulation were used in both cases, including a transient response. $63 \%$ of the time series was used as training data, based on which the reduced-order model is constructed. The coupled system of equations \eqref{flowROM} and \eqref{CrNic} was then simulated and VIV predictions beyond the training time were examined. 

\subsection{Velocity flowfield POD Basis}
\label{basis}

Before presenting the ROM predictions, it is interesting to examine the POD basis onto which the data-driven model for the velocity flowfield is projected (see \Cref{proj}). We will hereby focus on selected modes $\Phi_i$ for the two examined cases, which correspond to three significant phenomena in VIV dynamics. This analysis will help to interpret the simulation results obtained with our approach in \Cref{Vibr,Flowfield}, as well as provide insight into the projection step of \Cref{proj} and the main mechanisms of VIV dynamics.

For a detailed study focused on the POD of experimental VIV data, the reader is directed to \cite{Riches2018}.

\begin{figure}[!htbp]

\subfloat[$\Phi_1$ for $u_x$ velocity ($Re=90$)]{%
  \includegraphics[clip,width=1.1\columnwidth]{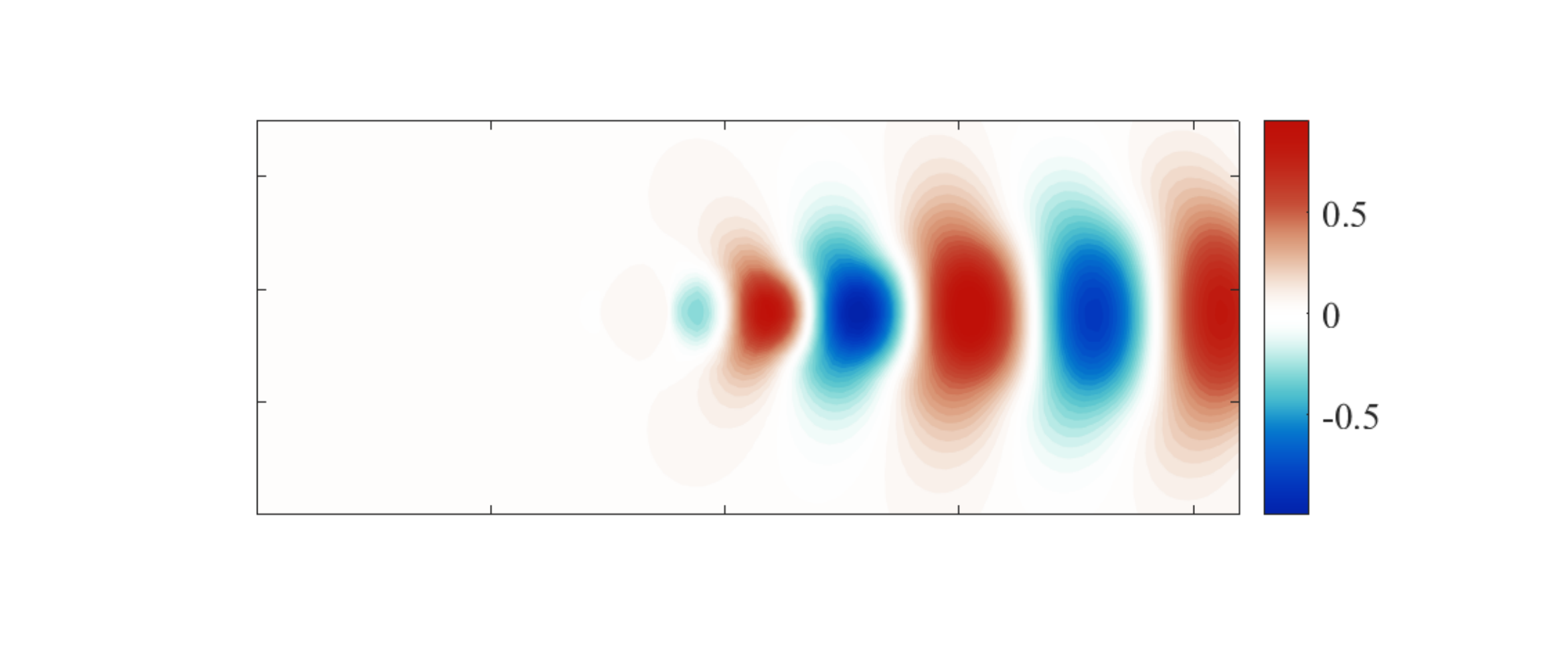}%
}

\subfloat[$\Phi_1$ for $u_y$ velocity, ($Re=90$)]{%
  \includegraphics[clip,width=1.1\columnwidth]{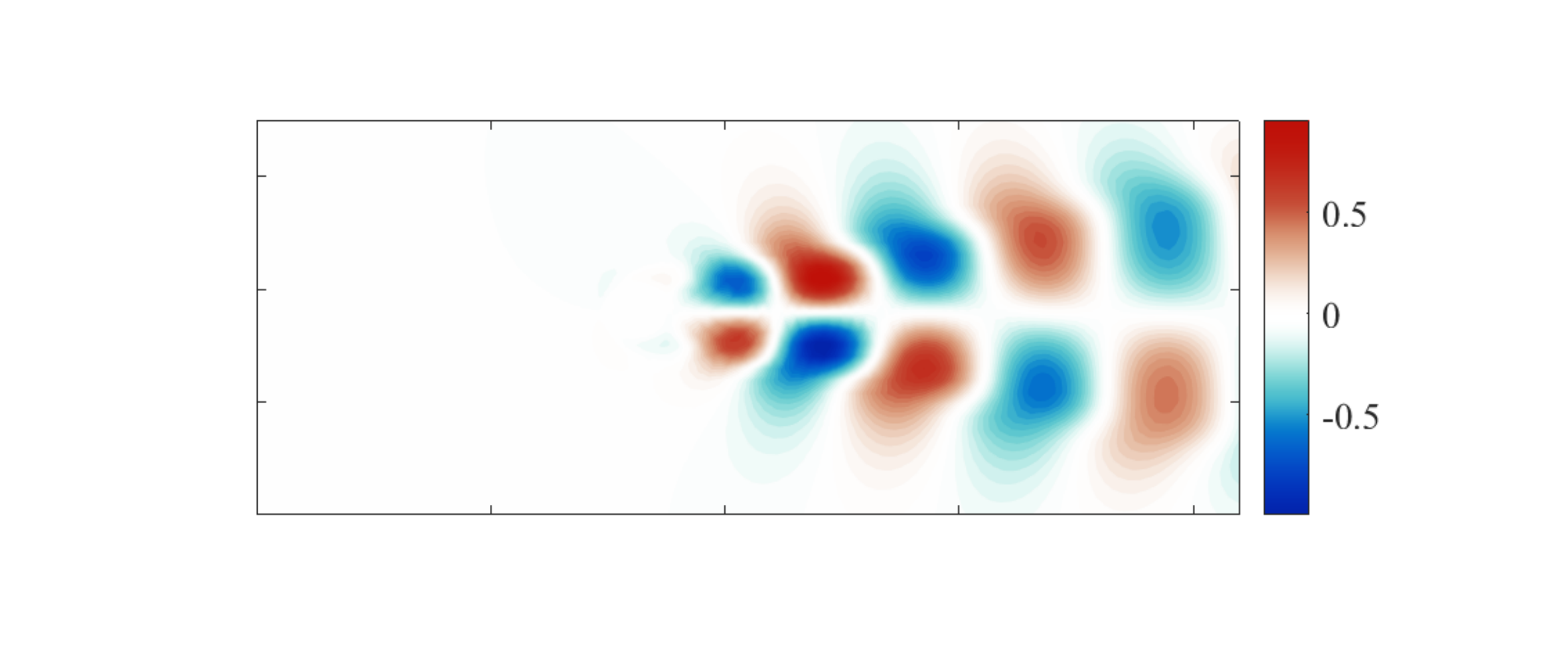}%
}

\subfloat[Fourier spectrum of $\Psi_1$ ($Re=90$)]{%
  \includegraphics[clip,width=1.1\columnwidth]{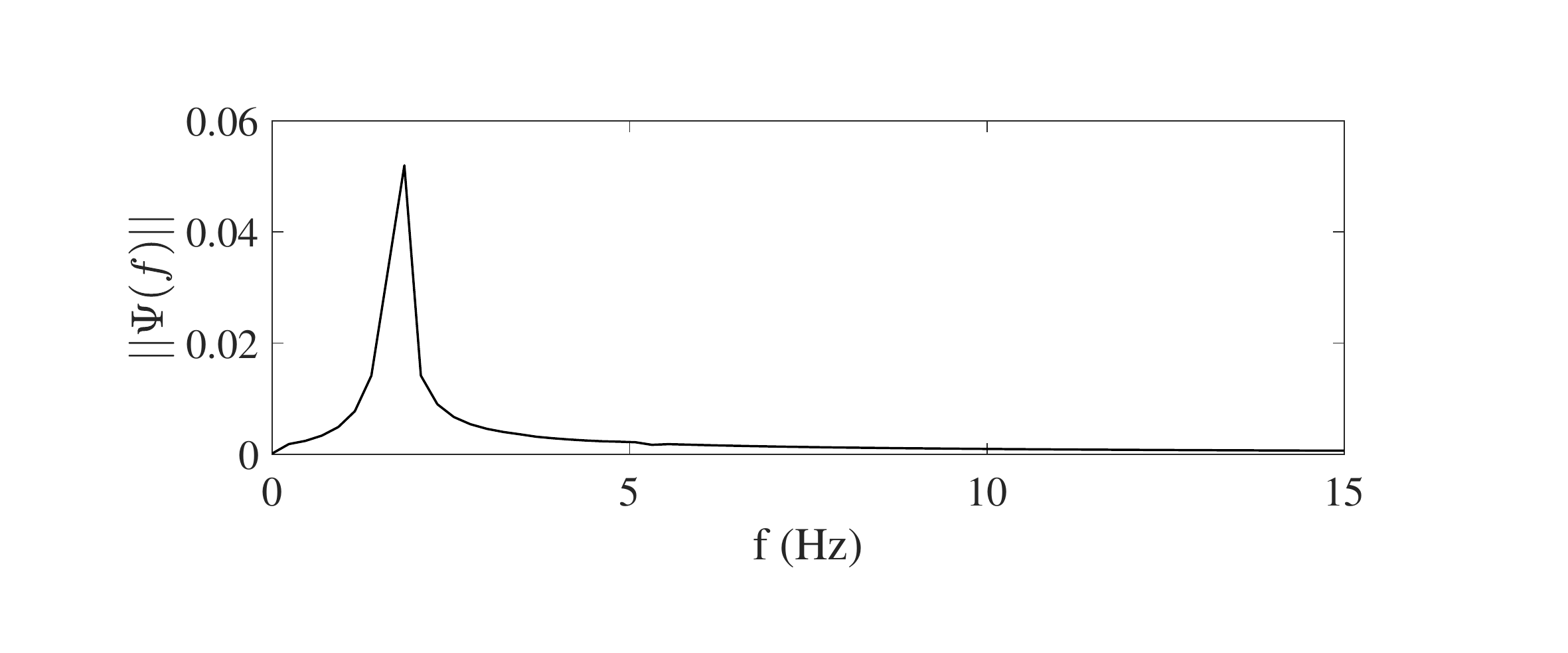}%
}

\caption{First POD mode for $Re=90$: The first POD mode pair contains more than $50\;\%$ of the total kinetic energy of the flow (here only the first mode is displayed) and is linked with the 2S vortex shedding frequency (peak at $f=1.72\; Hz$). The values of $\Phi_1$ are normalized to a $[-1,1]$ scale for illustration purposes.}
\label{fig:1stmode}
\end{figure}

After removing the data average (see \Cref{subsec:avg}), the first POD mode pair corresponds to convective vortex shedding for both $Re=90$ and $Re=180$. This is similar to the case of the flow over a cylinder \cite{Noack2003}, with this mode pair containing more than $50\;\%$ of the total kinetic energy of the flow. We hereby show only the first mode for $Re=90$, $u_x$ and $u_y$ (\Cref{fig:1stmode}a,b), since the second mode is just shifted by one quarter of the spatial wavelength. For a Strouhal number of $St=0.21$ and choosing a reference velocity as $u_{max}=1.13\;m/s$, the expected vortex shedding frequency is $f_v\approx1.97\; Hz$. This is in good accordance with the peak at $1.72\;Hz$ of the $\Psi_1$ Fourier spectrum, given in \Cref{fig:1stmode}c. 

\begin{figure}[htbp]

\subfloat[$\Phi_5$ for $u_x$ velocity ($Re=90$)]{%
  \includegraphics[clip,width=1.1\columnwidth]{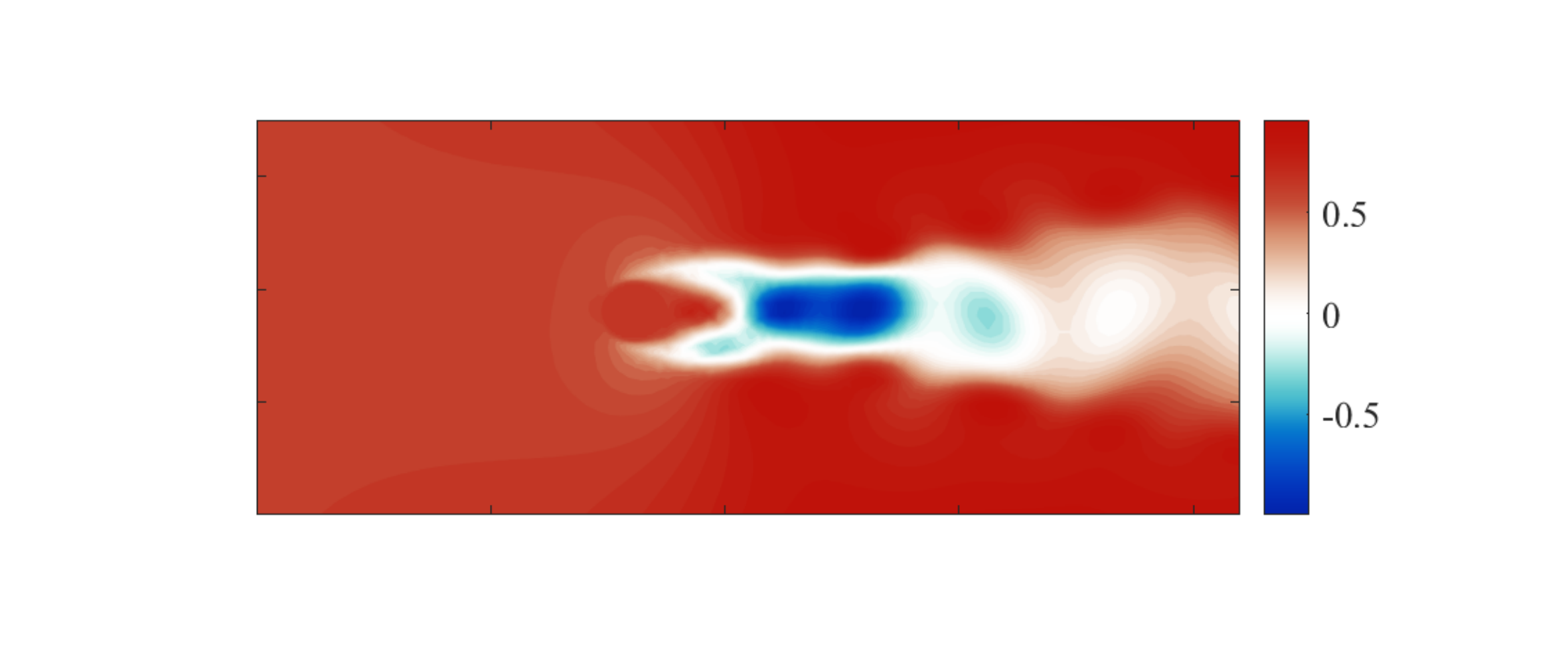}%
}

\subfloat[$\Phi_5$ for $u_y$ velocity ($Re=90$)]{%
  \includegraphics[clip,width=1.1\columnwidth]{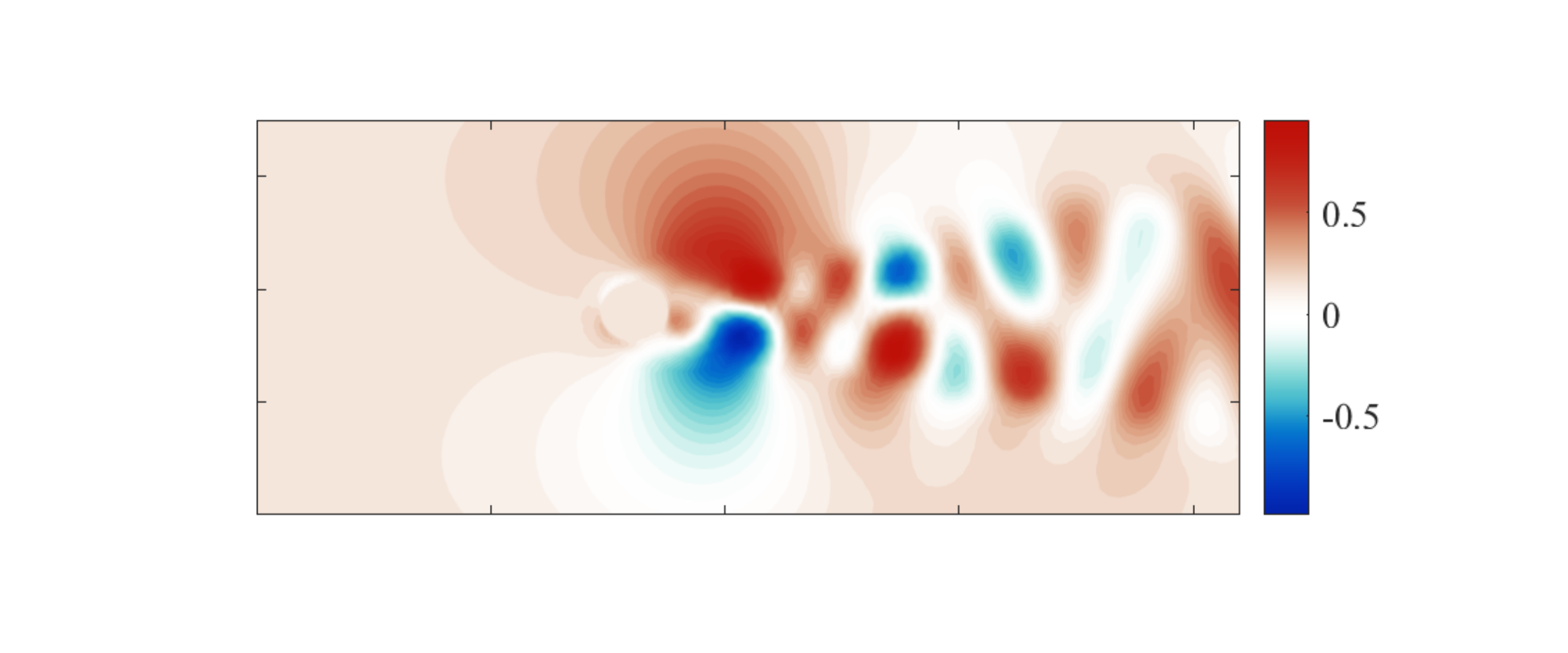}%
}

\subfloat[Fourier spectrum of $\Psi_5$ ($Re=90$)]{%
  \includegraphics[clip,width=1.1\columnwidth]{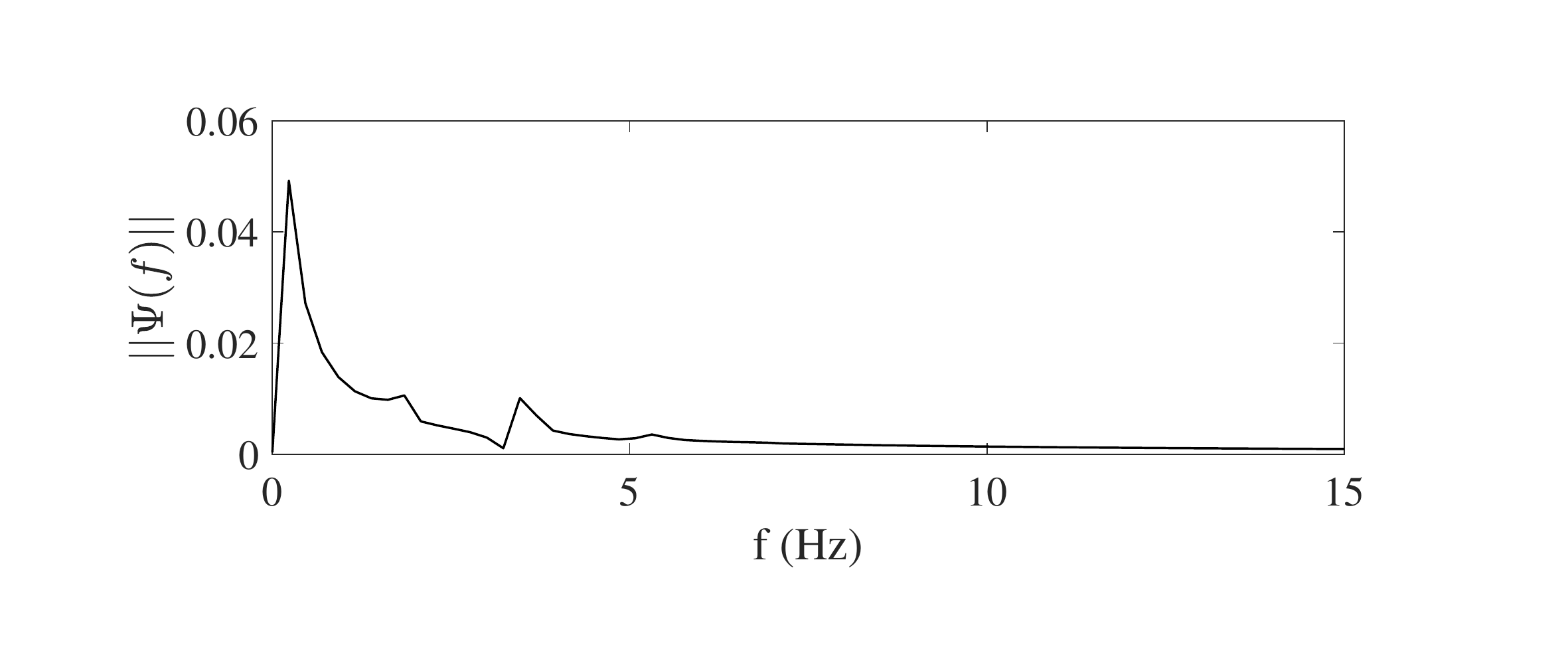}%
}

\caption{Fifth POD mode for $Re=90$: A slow-drift of the mean flow is captured, linked to the transient wake dynamics. Minor peaks at $f_v$ and $f_N$, possibly also due to transient phenomena.}
\label{fig:2ndmode}
\end{figure}

A mode indicatory of the slow-drift of the mean flow is also detected in both datasets. This mode ($\Phi_5$ for $Re=90$) captures the energy of flow transition to a limit-cycle behaviour and is illustrated in \Cref{fig:2ndmode} for both velocity components. It is clear that the significance of such a mode depends on the presence of transient dynamics in the VIV simulation dataset and in many cases is individually treated \cite{Riches2018}. The involved low-frequency dynamics are indicated by the corresponding Fourier spectrum in \Cref{fig:2ndmode}c, while two minor peaks are noticed close to the vortex shedding and solid natural frequencies.

\begin{figure}[htbp]

\subfloat[$\Phi_5$ for $u_x$ velocity ($Re=180$)]{%
  \includegraphics[clip,width=1.1\columnwidth]{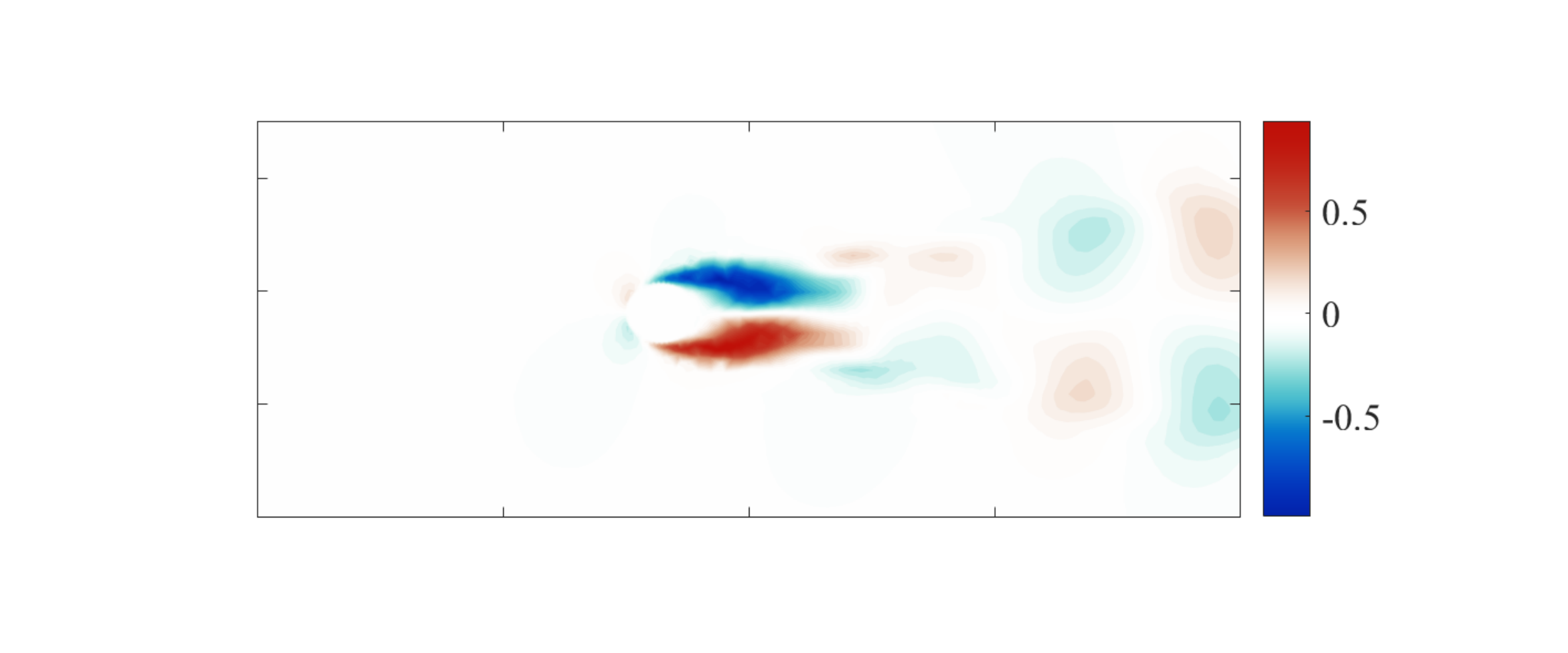}%
}

\subfloat[$\Phi_5$ for $u_y$ velocity ($Re=180$)]{%
  \includegraphics[clip,width=1.1\columnwidth]{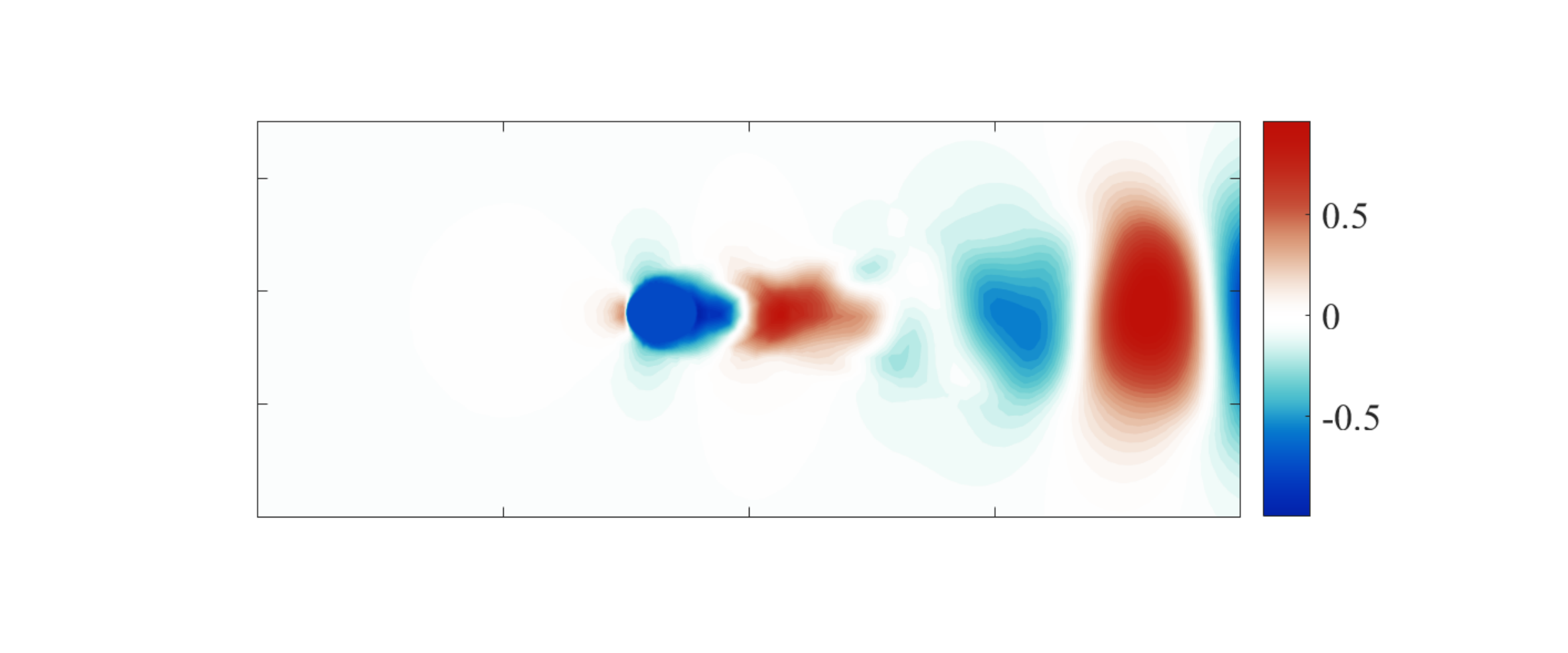}%
}

\subfloat[Fourier spectrum of $\Psi_5$ ($Re=180$)]{%
  \includegraphics[clip,width=1.1\columnwidth]{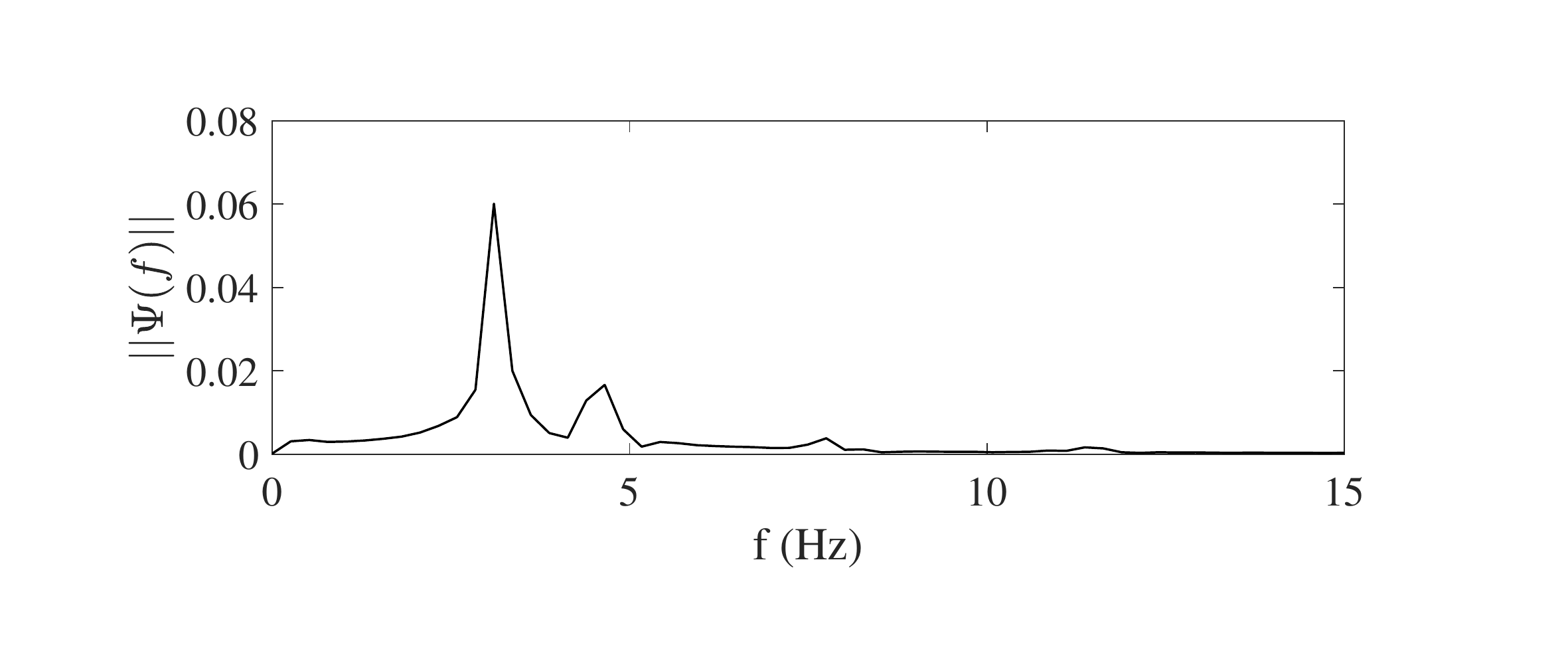}%
}

\caption{Fifth POD mode for $Re=180$: Solid natural frequency affects the wake dynamics, since the case lies in the right side of the initial VIV branch. A significant peak at $3.12\; Hz$, very close to the theoretical $f_N$ value is observed in the Fourier spectrum of $\Psi_5$.}
\label{fig:3rdmode}
\end{figure}

In the case of $Re=90$, other POD modes correspond mainly to multiples of the vortex-shedding frequency and corresponding wavelengths, since the solid forcing frequency is sufficiently below $f_N=3.24\; Hz$. 
However, in the case of $Re=180$, the two frequencies are very close. This is specifically evident in $\Phi_5$ of the corresponding basis, shown in \Cref{fig:3rdmode}. The Fourier transform of $\Psi_5$ (\Cref{fig:3rdmode}c) shows a strong peak at $3.12\; Hz$, indicating the presence of the solid natural frequency in the system dynamics. This is expected theoretically \cite{Williamson2004} for $Re=180$ and can be captured by the ROM with more than $5$ modes. It is noted that a departure from the exact value of $f_v/f_N=1$ can be expected in cases of low mass ratio and damping \cite{Khalak1999}, as the one considered here.

\subsection{Solid Oscillations}
\label{Vibr}

\begin{figure}[!htbp]
\subfloat[Predictions for streamwise oscillation velocity, $Re=90$. ]{%
  \includegraphics[clip,width=1.1\columnwidth]{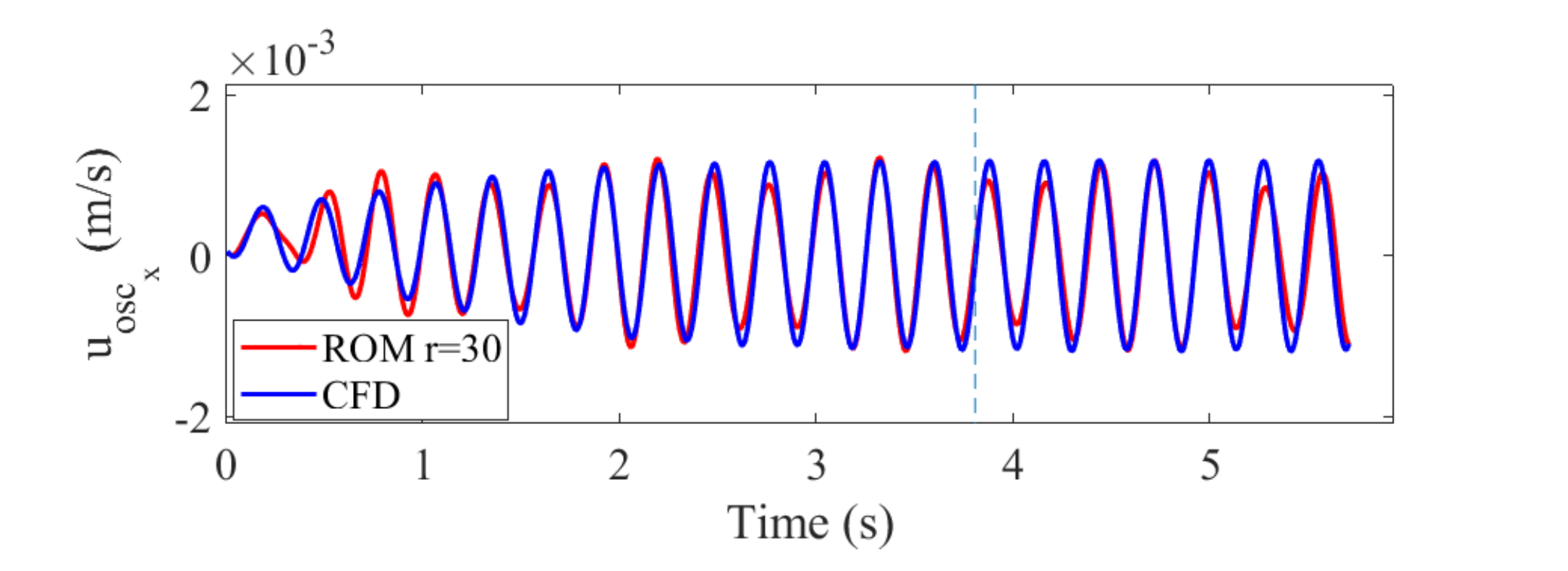}%
}

\subfloat[Predictions for transverse oscillation velocity, $Re=90$.]{%
  \includegraphics[clip,width=1.1\columnwidth]{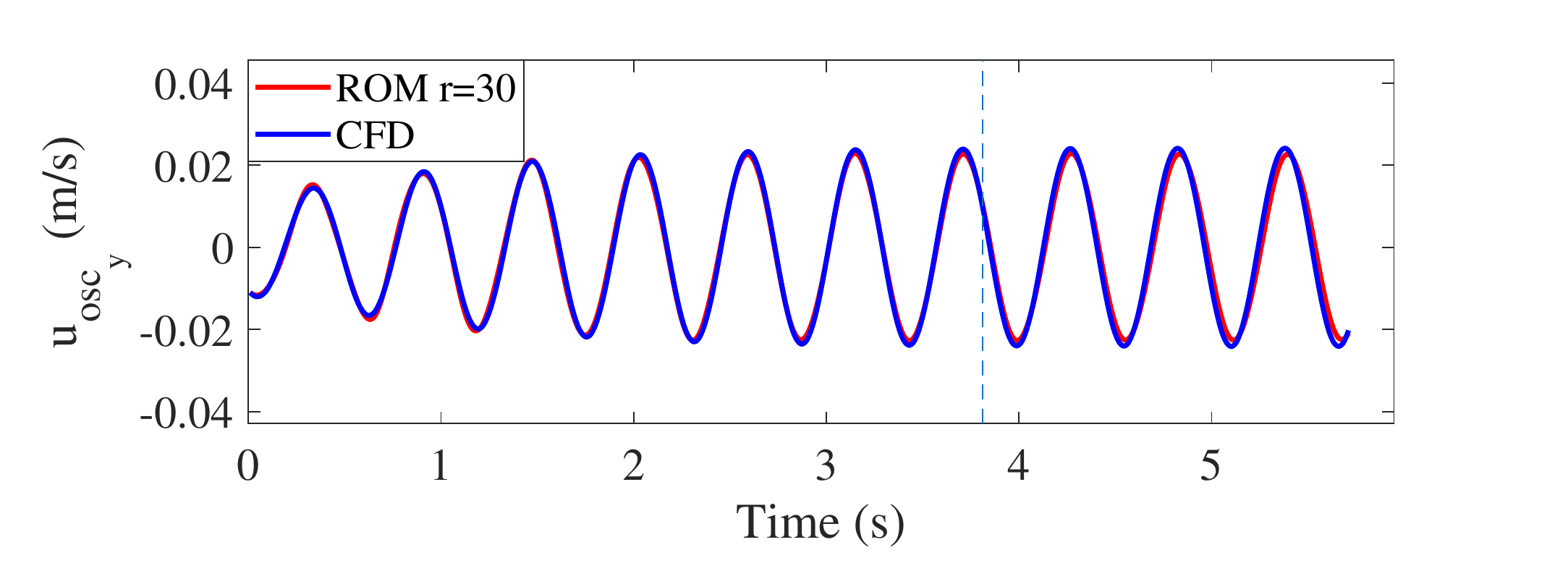}%
}

\subfloat[Predictions for streamwise oscillation velocity, $Re=180$.]{%
  \includegraphics[clip,width=1.1\columnwidth]{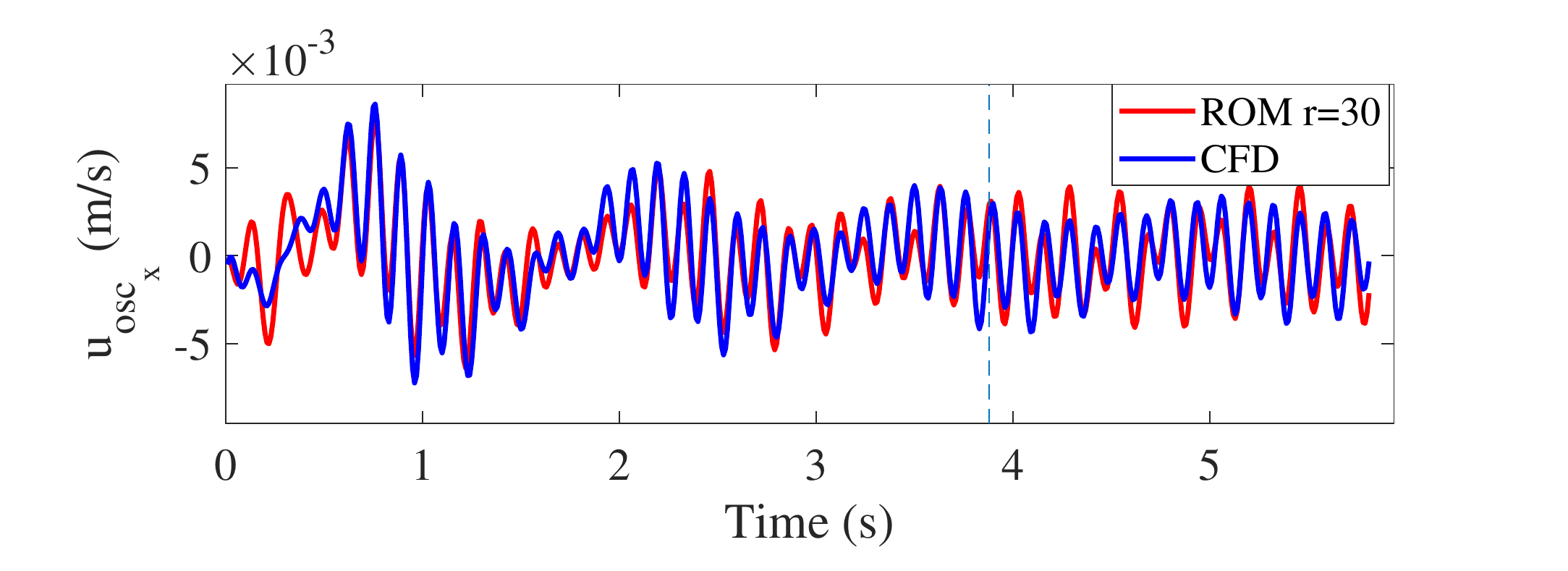}%
}

\subfloat[Predictions for transverse oscillation velocity, $Re=180$.]{%
  \includegraphics[clip,width=1.1\columnwidth]{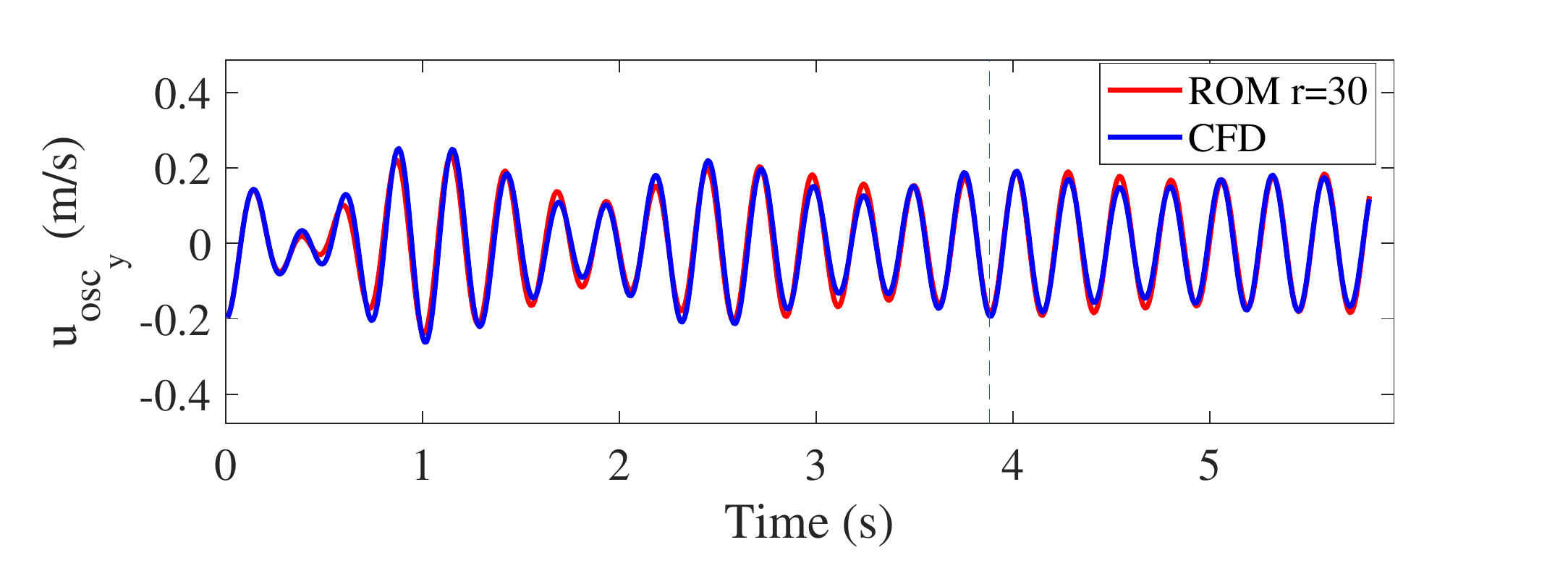}%
}
\caption{Solid oscillations predictions: The coupled dynamics of the data-driven flow ROM \eqref{flowROM} and first-principle oscillation equations \eqref{CrNic} yield highly accurate predictions for the solid oscillations, capturing well the low-amplitude motion ($Re=90$) as well as beating phenomena ($Re=180$), even past the training time.}
\label{fig:osc_velo}
\end{figure}

The oscillation velocity time series and corresponding ROM predictions are given in \Cref{fig:osc_velo} for both testcases. The vertical dashed line indicates the end of the training time. From that point and on, the model predictions are tested against unseen simulation data. In both cases, the streamwise oscillation amplitude (along $x$) is more than one order of magnitude lower compared to the amplitude of the transverse oscillation. Furthermore, the transverse oscillations dominating frequency ($f\approx 1.72 \; Hz$ for $Re=90$) is half than that of streamwise motion, as theoretically expected.

For the case of $Re=90$, the oscillations exhibit a very low-amplitude (\Cref{fig:osc_velo}b), hinting a link to the lower end of the initial VIV branch \cite{Navrose2014}. On the contrary, the $Re=180$ case shows strong transverse oscillations (\Cref{fig:osc_velo}d), almost one order of magnitude higher than that for $Re=90$, often exhibited in the higher end of the initial VIV branch. Due to the vicinity of $f_N$ and $f_v$, beating phenomena are observed for $Re=180$  \cite{Shen2018}. The coupled data-driven-oscillation model captures the solid dynamics with high accuracy in both cases.

The qualitative differences in the transverse and streamwise oscillations are reflected on the ROM predictions error. The truncation of the SVD modes \eqref{phi} leads to a corresponding truncation of the flow energy. If the VIV coupling is strong (as in the streamwise direction), even leading flow modes interact with the solid oscillation (see e.g. \Cref{fig:3rdmode}). However, in cases of weaker VIV coupling (as in \Cref{fig:osc_velo}a,c), it is likely that low energy modes that interact with the low-amplitude solid motion are truncated. Hence, the relative error for the $r=30$ ROM prediction e.g. in \Cref{fig:osc_velo}c might seem significant at specific times, however the absolute ROM prediction error is negligible.

\subsection{Flowfield Predictions}
\label{Flowfield}

For each simulation timestep, we can reconstruct the predicted velocity field in the reference configuration, by $\mathbf{u}_r=\Phi_r^T \Phi_r \tilde{\mathbf{u}}$. However, it is of interest to also map the reconstructed flowfield to the predicted current configuration. This can be done by using the dimensionless deformation field $\left(\sum_{i \in I}{\Lambda^{-1}_{:,i}}\right)$ from \eqref{disc_Lap} and multiplying with the predicted displacement of the solid body along both axes. The solution is then mapped to node positions of the predicted deformed grid. This last step entrails a negligible added cost, since $\sum_{i \in I}{\Lambda^{-1}_{:,i}}$ has been precomputed. 

\begin{figure}[!htbp]

\subfloat[CFD contour plot for $u_x$ ($Re=90$)]{%
  \includegraphics[clip,width=1.1\columnwidth]{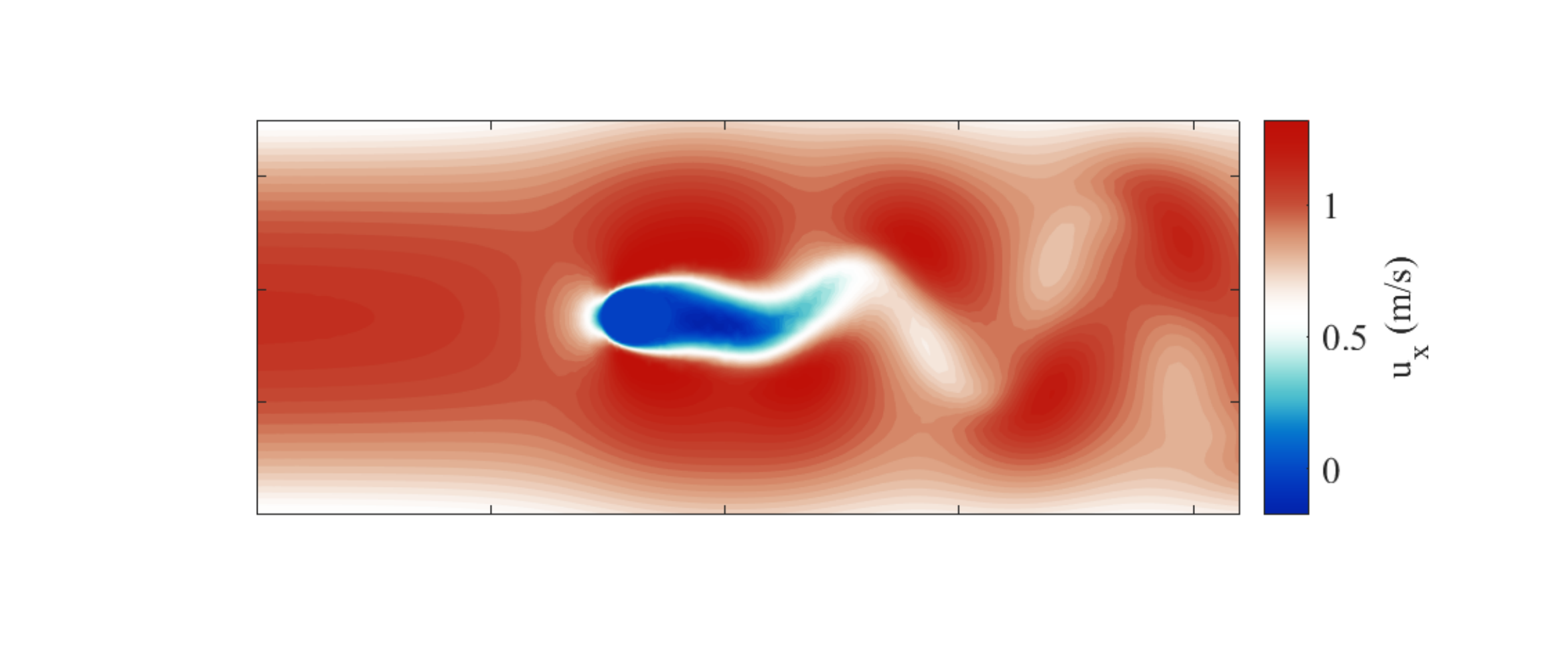}%
}

\subfloat[ROM ($r=30$) contour plot for $u_x$ ($Re=90$)]{%
  \includegraphics[clip,width=1.1\columnwidth]{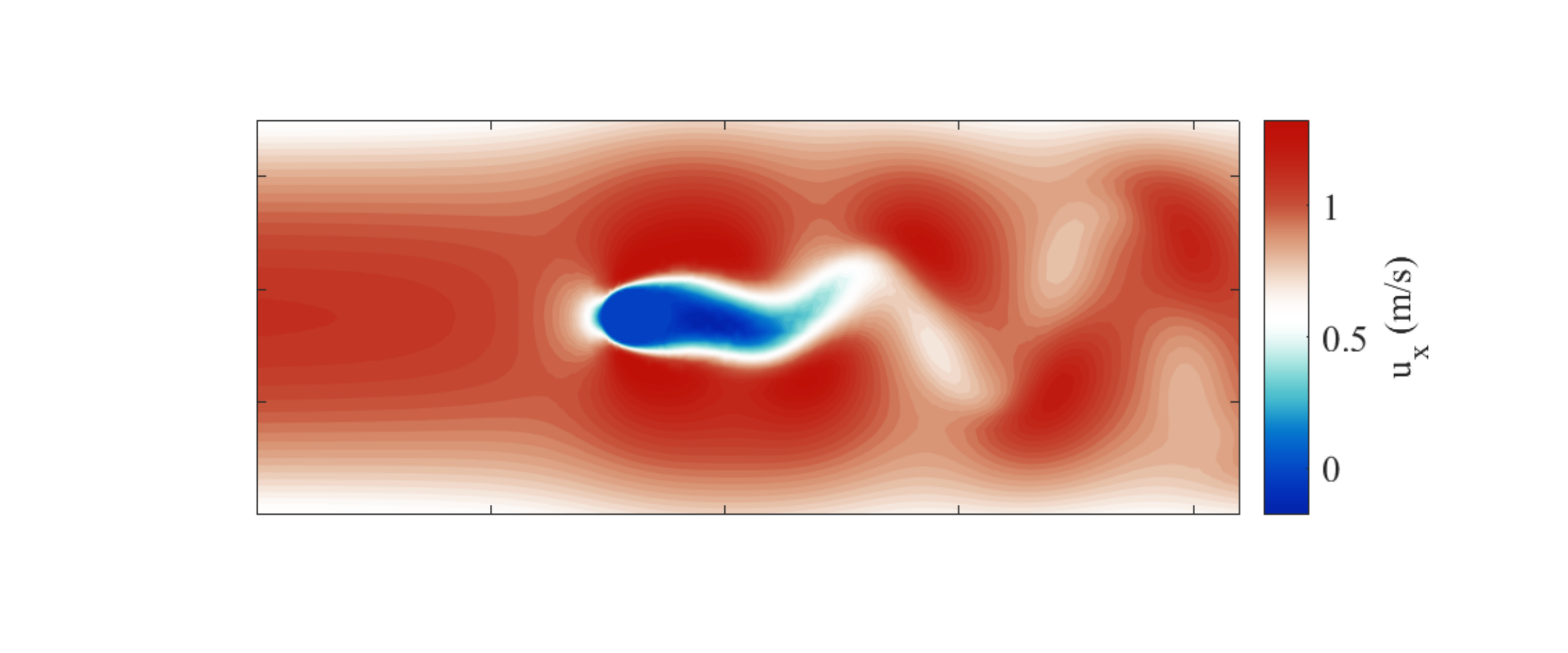}%
}
\caption{Contour plots at $t=5.9 s$ for the $u_x$ flowfield at $Re=90$: The non-intrusive model captures the 2S vortical mode as well as the flow features close to the oscillating body.}
\label{fig:field_u_comp90}
\end{figure}

\begin{figure}[!htbp]
\subfloat[CFD contour plot for $u_y$ ($Re=90$)]{%
  \includegraphics[clip,width=1.1\columnwidth]{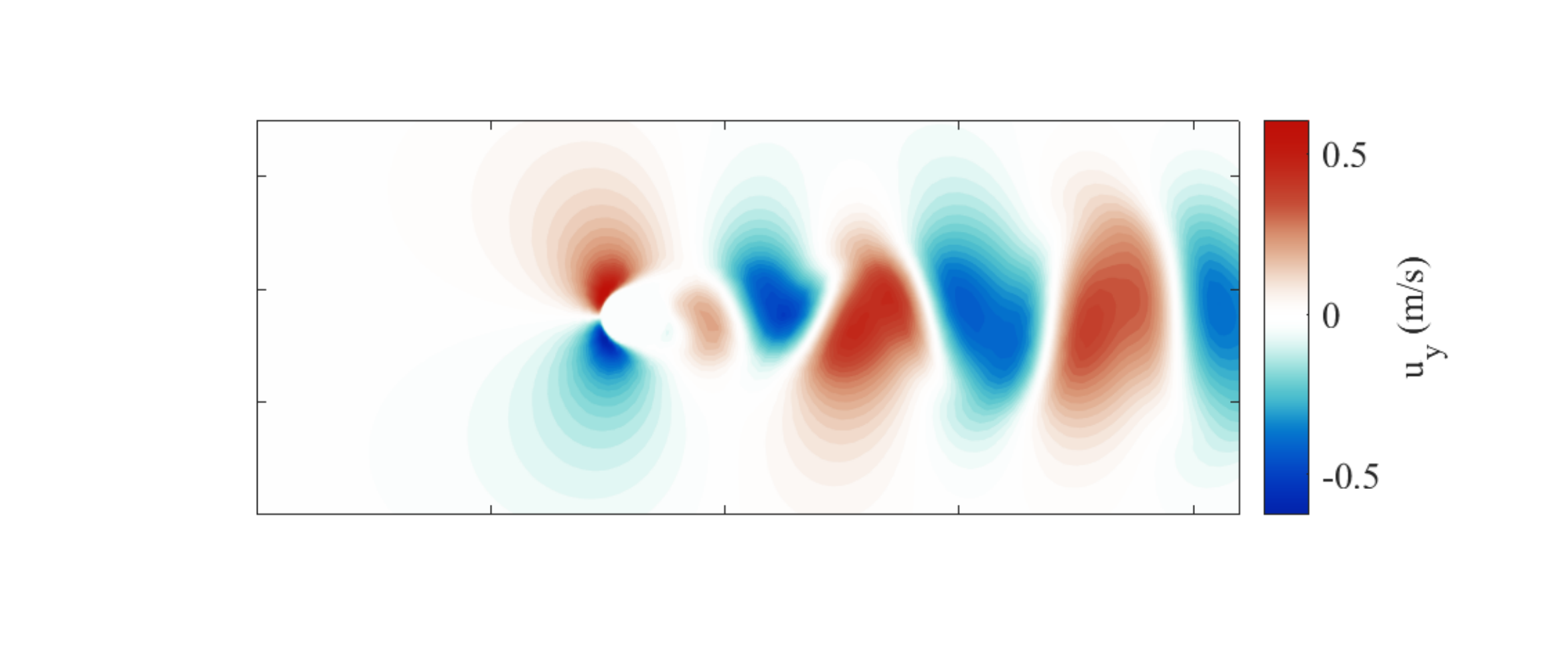}%
}

\subfloat[ROM ($r=30$) contour plot for $u_y$ ($Re=90$)]{%
  \includegraphics[clip,width=1.1\columnwidth]{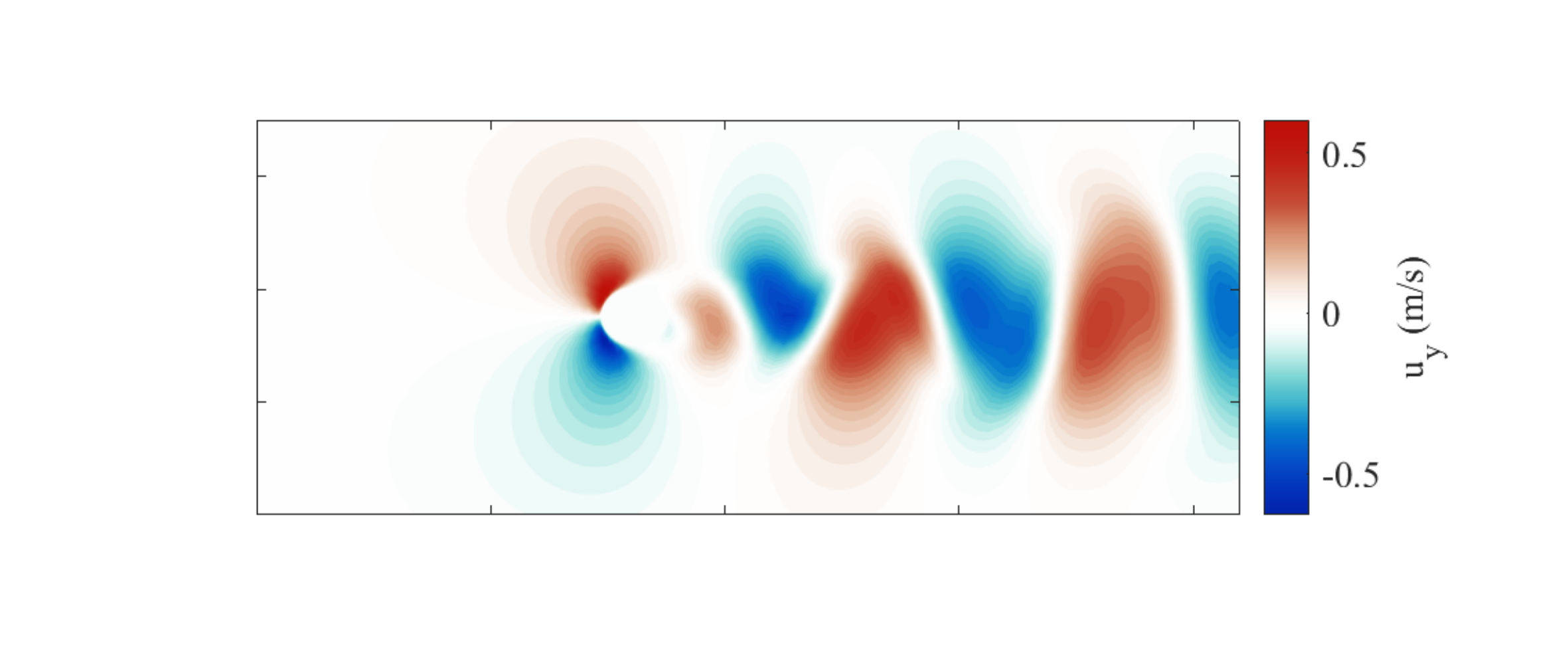}%
}
\caption {Contour plots at $t=5.9 s$ for the $u_y$ flowfield at $Re=90$: Slight phase mismatch, also indicated by the solid transverse oscillation prediction in \Cref{fig:osc_velo}b.}
\label{fig:field_v_comp90}
\end{figure}

\begin{figure}[!htbp]

\subfloat[CFD contour plot for $u_x$ ($Re=180$)]{%
  \includegraphics[clip,width=1.1\columnwidth]{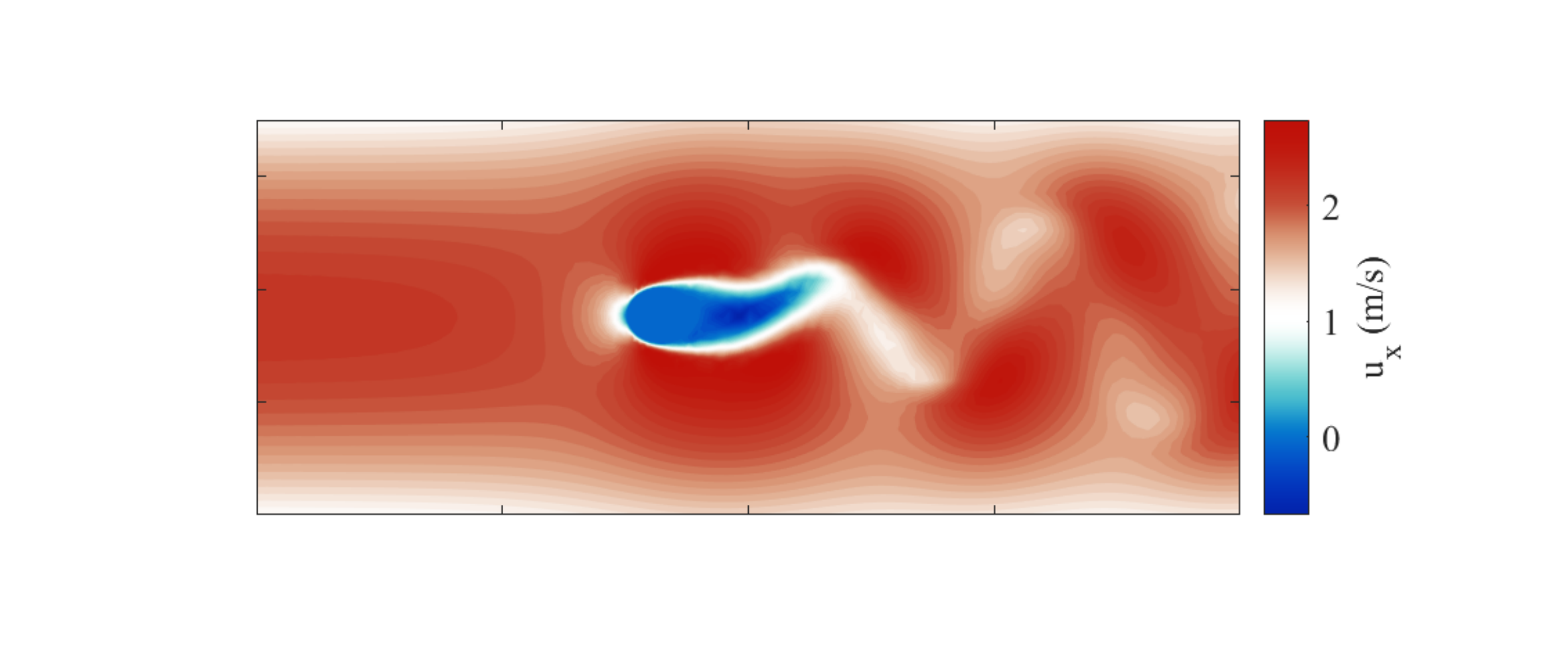}%
}

\subfloat[ROM ($r=30$) contour plot for $u_x$ ($Re=180$)]{%
  \includegraphics[clip,width=1.1\columnwidth]{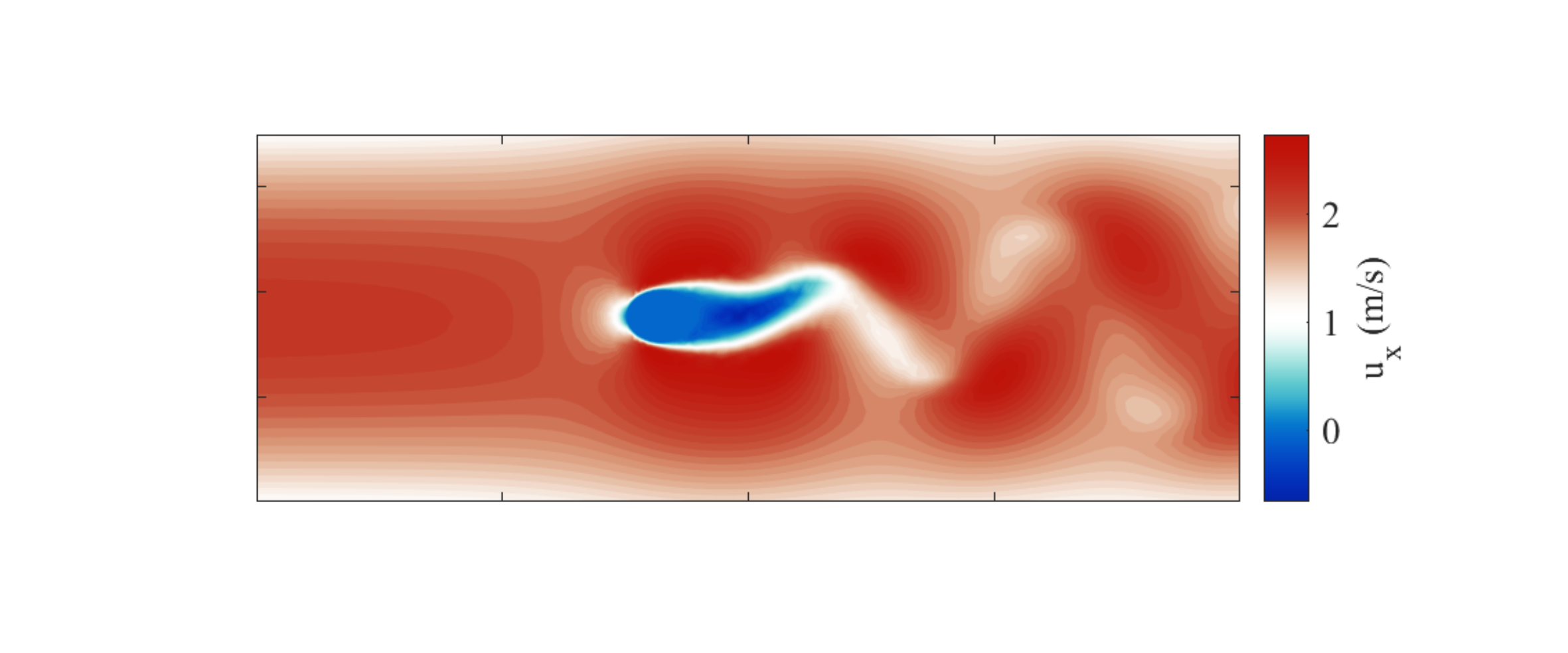}%
}
\caption{Contour plots at $t=5.9 s$ for the $u_x$ flowfield at $Re=180$: Accurate ROM prediction, with a double vortex shedding frequency compared to $Re=90$.}
\label{fig:field_u_comp180}
\end{figure}

\begin{figure}[!htbp]
\subfloat[CFD contour plot for $u_y$ ($Re=180$)]{%
  \includegraphics[clip,width=1.1\columnwidth]{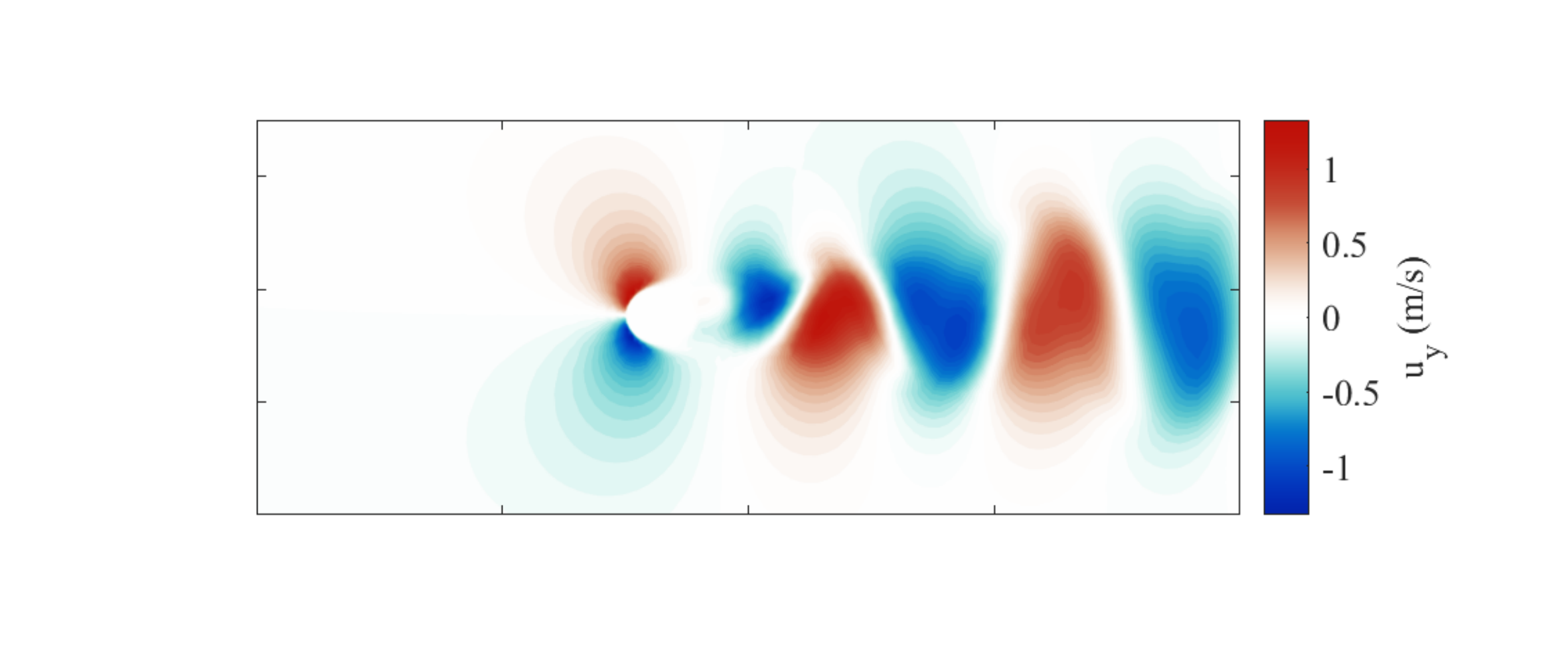}%
}

\subfloat[ROM ($r=30$) contour plot for $u_y$ ($Re=180$)]{%
  \includegraphics[clip,width=1.1\columnwidth]{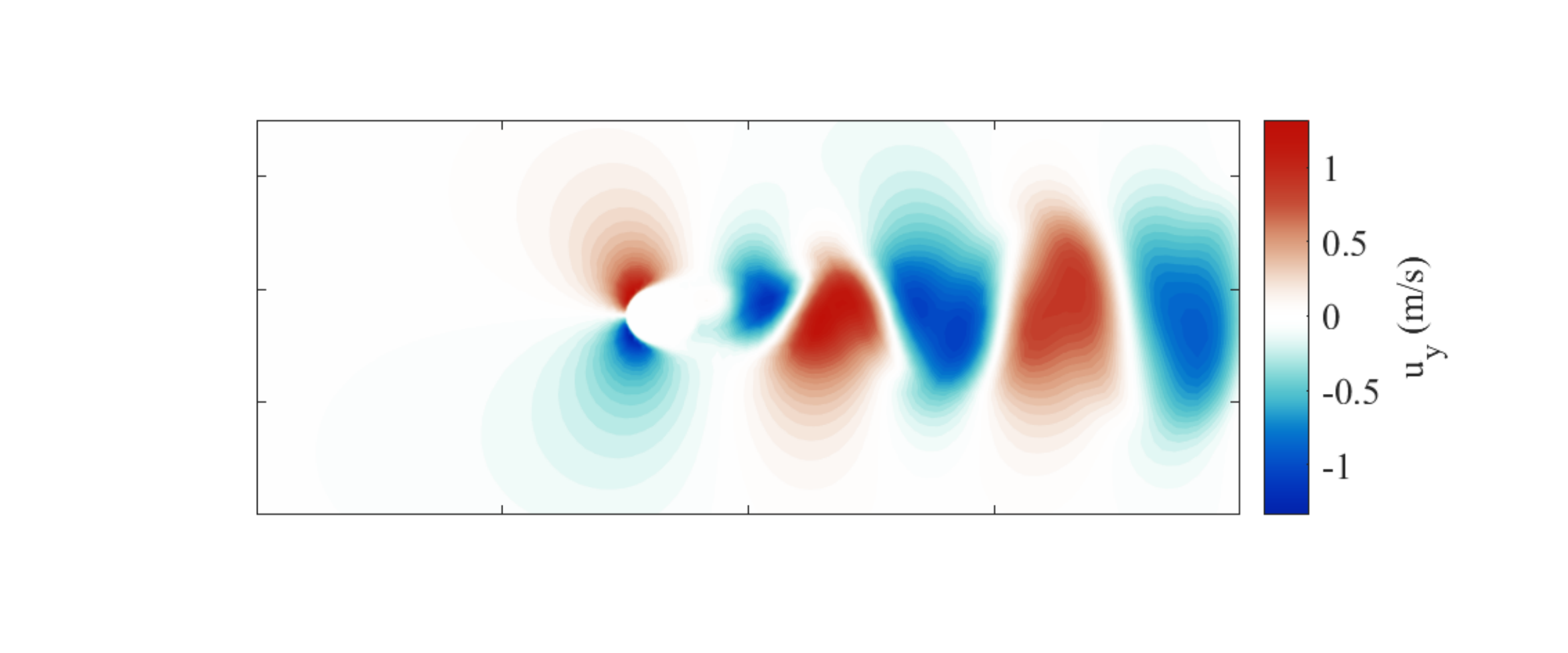}%
}
\caption{Contour plots at $t=5.9 s$ for the $u_y$ flowfield at $Re=180$: The non-intrusive model is again in good accordance with the corresponding CFD results at the end of testing time.}
\label{fig:field_v_comp180}
\end{figure}

The CFD and predicted ROM flowfields for the final testing time are given in \Cref{fig:field_u_comp90} and \Cref{fig:field_v_comp90}, for both velocity components of $Re=90$. A corresponding comparison is given for $Re=180$ in \Cref{fig:field_u_comp180} and \Cref{fig:field_v_comp180}. The CFD solution indicates a 2S mode for both testcases, where two single vortices appear per oscillation cycle. This matches well-known results for laminar VIV with free body oscillations \cite{Williamson2004, Navrose2014}. ROM predictions in both cases match the vortex dynamics qualitatively and quantitatively, employing $r=30$ DOFs. It is noted that the full-order numerical simulation (e.g. for $Re=180$) requires approximately $40$ minutes on a personal laptop. In contrast, the offline ROM cost corresponds to approximately $3$ minutes, while the ROM numerical simulation requires just several seconds (for some typical ROM dimension of $\mathcal{O}(10)$).

%Re90

%Re180

We also compute an average, relative ROM error over the $u_x$ and $u_y$ flowfield. This reads (e.g. for the $u_x$ component) as

\begin{equation}
\label{averror}
  e=\frac{{\|{\mathbf{u}_{{x}_{CFD}}-\mathbf{u}_{x_{ROM}}}\|}_1} {n\;\max{\mathbf{u}_{{x}_{CFD}}}}\times 100\%.
\end{equation}

\begin{figure}[!htbp] 
  \begin{center}
    \includegraphics[clip,width=1.05\columnwidth]{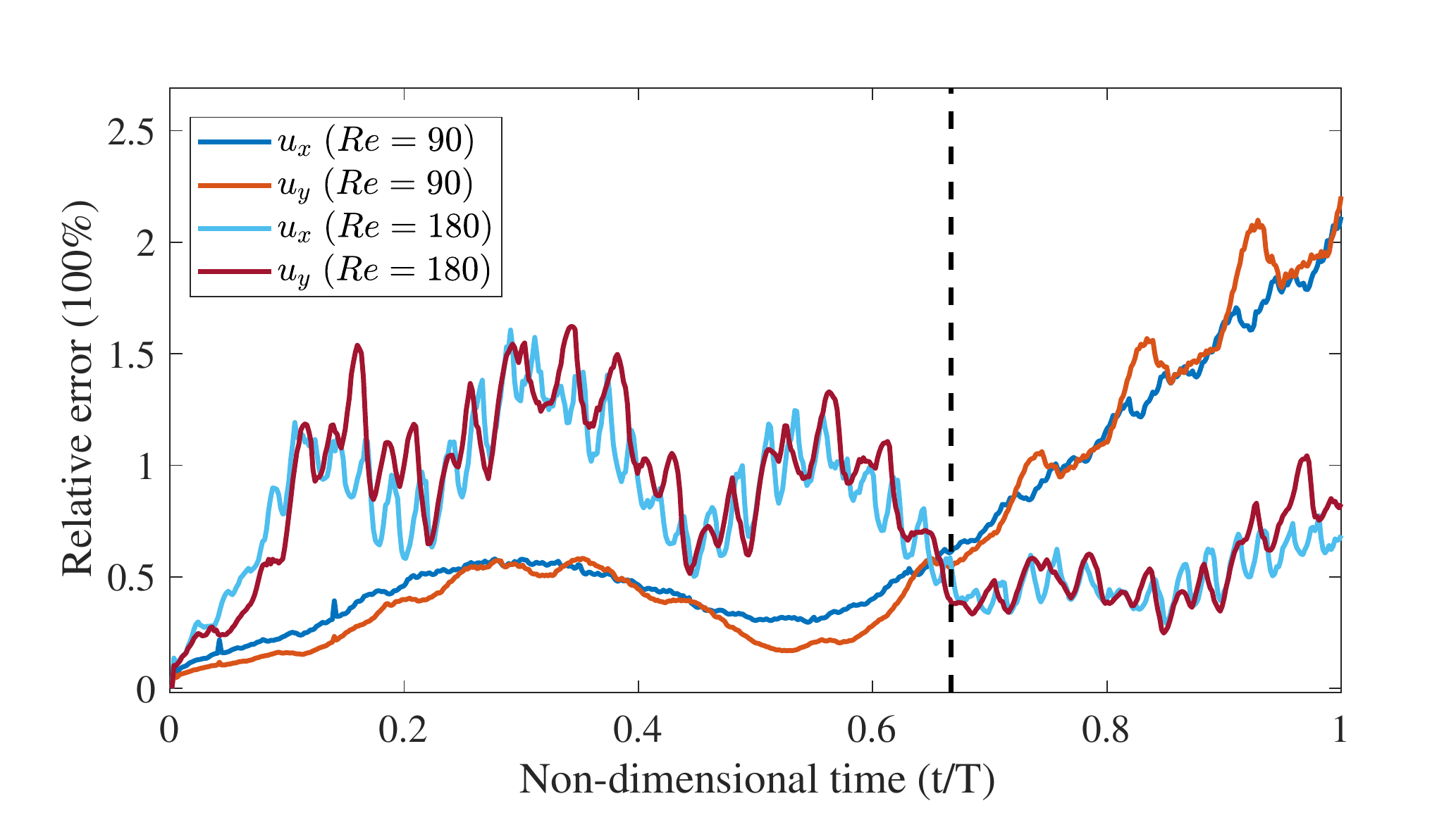}
    \caption{Average error over the flowfield, for both $u_x$ and $u_y$ relative to $\max{u_x}$, $\max{u_y}$ over time.}
    \label{fig:errors}
  \end{center}
\end{figure}

Error $e$ is recorded in \Cref{fig:errors} over time, for both testcases and both velocity components. For $Re=180$, the error exhibits oscillations, possibly linked to both modeling and projection errors of the stronger vortex dynamics, compared to $Re=90$. We observe that in the transient part of the data, the error reaches its peak value, approximately $1.5 \%$. It is noted that the flowfield error originates from the combined effect of modelling and projection errors, as well as an error from the coupling of \eqref{flowROM} with \eqref{CrNic}. In the testing time interval, the errors for $Re=180$ (both $u_x$ and $u_y$) oscillate for values below $1 \%$. For $Re=90$, the error shows a noticeable increase after the training time, reaching up to $2 \%$ at the end of the testing time. This is primarily linked with a phase shift, slightly noticeable in \Cref{fig:osc_velo}a. Transient dynamics induced by the slow, mean flow drift are not fully resolved during the training time (see the low-frequency peak in \Cref{fig:3rdmode}c). Thus, a gradual phase shift of the ROM predictions is observed during the testing time interval, leading to a respective, almost linear increase of the error in \Cref{fig:errors}.

\begin{figure}[!htbp] 
  \begin{center}
    \includegraphics[clip,width=1.05\columnwidth]{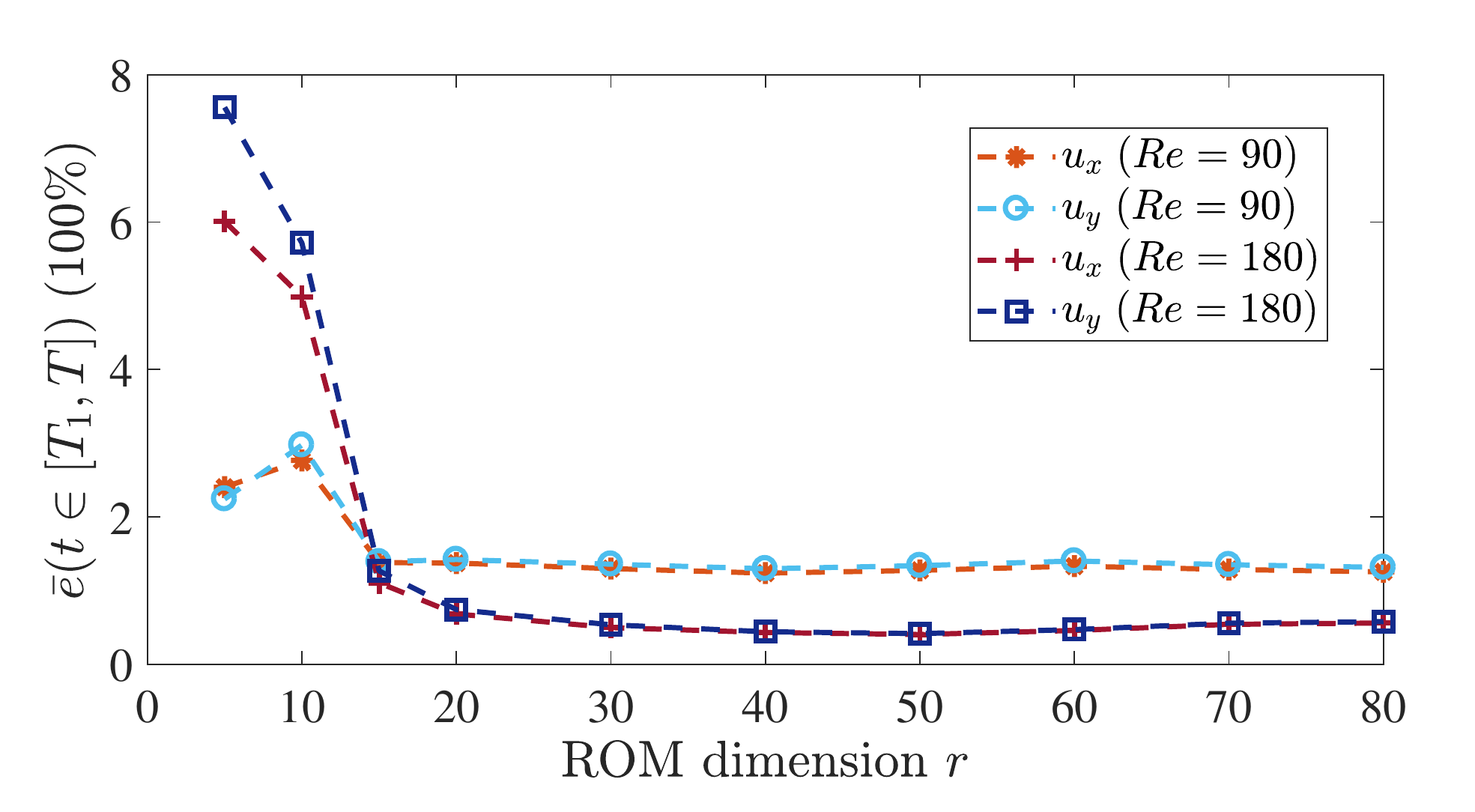}
    \caption{Average error \eqref{averror} over testing time, for different ROM dimensions $r$: A projection subspace dimension of $[15, 20]$ is sufficient for the ROM, with an observed error convergence for a further $r$ increase.}
    \label{fig:ROMdim}
  \end{center}
\end{figure}

Finally, we perform a comparative study by averaging the error \eqref{averror} over the testing time, with respect to the dimension $r$ of the ROM. As presented in \Cref{adj} and \Cref{proj}, the full-order data-driven flow model is independent of the projection basis $\Phi_r$ (eq. \eqref{phi}). After truncating the projection basis to 10 different $r$ values, we simulate the resulting coupled system and record the average of \eqref{averror} from the end of training time $T_1$ to the final simulation time $T$. \Cref{fig:ROMdim} illustrates the obtained results for both testcases and both velocity components. In the case of $Re=180$, the resulting $u_y$ error is slightly higher than $u_x$, especially for $r\leq 10$. For small values of $r$, the average error is higher for the $Re=180$ case. That can be explained by the more rich dynamics exhibited for this case, as shown in \Cref{fig:3rdmode} and \Cref{fig:osc_velo}c,d. Thus, comparably more POD modes are required to sufficiently capture the system dynamics. In both cases, there appears a sharp corner for a ROM dimension close to $20$, meaning that adding further POD modes does not provide any further information on the model dynamics \cite{Benner2021}. However, \Cref{fig:ROMdim} also indicates a convergence of the ROM error to a certain value with increasing ROM dimension $r$. This property is typically acquainted in intrusive MOR (e.g. \cite{Lieu2005, Nonino2021}), here inherited by the presented non-intrusive ROM methodology. A similar error convergence behaviour with the one observed for the non-intrusive model in \Cref{fig:ROMdim} was reported for an intrusive FSI ROM in \cite{Nonino2021}. This indicates that the independence of the inferred sparse, full-order operators from the projection basis $\Phi_r$ could result to an increased robustness of the non-intrusive ROM.

\section{Outlook and Future Work}%
\label{sec:con}

In this work, a non-intrusive model order reduction methodology was presented, applied to vortex-induced vibration problems. The dynamical model structure for the fluid velocity was motivated by the problem physics under the ALE formulation, while a physics-based sparsity pattern was enforced through the grid adjacency information. An $L_2$ regularization term was added to the corresponding least squares problem for each grid node velocity component and computational cost reduction strategies for the optimization of the regularization parameter were discussed. The data-driven velocity field ROM was subsequently coupled with the first-principle solid oscillations and a method for mapping the solution from the reference to the deformed configuration was presented. Finally, the methodology was applied to two transient VIV testcases with $Re=90$, $Re=180$, which exhibit different VIV phenomena. Results on the prediction of both the solid oscillation and the surrounding flowfield with only $30$ DOFs indicated the potential of the followed methodology; the solid oscillation was accurately predicted over the testing time interval, while the average velocity error over the domain was found to be less than $3 \%$. Finally, a parametric study with respect to the ROM dimension showcased the increased ROM robustness offered by the inference of sparse, full-order operators. 

The current work could comprise a first step towards a complete non-intrusive MOR framework for FSI problems. Hence, future work could incorporate several different aspects; in particular, the offline computational efficiency of the approach could be significantly enhanced through parallelizing least-squares problems and domain segregation. Moreover, testing the developed non-intrusive ROMs under varying inputs and solid oscillation parameters would provide further insight on the attributes of the proposed methodology. In a similar direction, the showcased sparse full-order model independence from the SVD basis renders parametric MOR specifically interesting, aiming towards non-smooth parameter dependencies. Such studies could be performed for VIV problems with respect to different Reynolds numbers, given the recorded rich coupled system dynamics and strong parametric dependence. Finally, the extension to deformable solids and consequently the formulation of a predictive, non-intrusive ROM methodology for FSI problems lies in the future scope of this work.

\FloatBarrier
%%%%%%%%%%%%%%%%%%%%%%%%%%%%%%%%%%%%%%%%%%%%%%%%%%%%%%%%%%%%%%%%%%%%%%%%%%%%%%%%
% *** REFERENCES ***                                                           %
%%%%%%%%%%%%%%%%%%%%%%%%%%%%%%%%%%%%%%%%%%%%%%%%%%%%%%%%%%%%%%%%%%%%%%%%%%%%%%%%
\addcontentsline{toc}{section}{References}
\bibliographystyle{plainurl}
\bibliography{main}
  
\end{document}